\documentstyle[12pt]{article}
\input epsf

\begin{document}
\baselineskip=15pt

\newcommand{\be}{\begin{equation}}
\newcommand{\ee}{\end{equation}}
\newcommand{\bq}{\begin{eqnarray}}
\newcommand{\eq}{\end{eqnarray}}
\newcommand{\x}{{\bf x}}
\newcommand{\y}{{\bf y}}

\newcommand{\p}{\varphi}
\newcommand{\Sc}{Schr\"odinger\,}
\newcommand{\del}{\nabla}
\begin{titlepage}
\rightline{DTP/99/41}
\vskip1in
\begin{center}
{\large One-loop Conformal Anomalies from AdS/CFT in the \Sc Representation.} 
\end{center}
\vskip1in
\begin{center}
{\large
Paul Mansfield and David Nolland

Department of Mathematical Sciences

University of Durham

South Road

Durham, DH1 3LE, England}

{\it P.R.W.Mansfield@durham.ac.uk}, {\it D.J.Nolland@durham.ac.uk}

\end{center}
\vskip1in
\begin{abstract}

\noindent 
We compute the conformal anomalies of boundary CFTs for
scalar 
and fermionic fields
propagating in AdS spacetime at one-loop. The coefficients are quantized, with
values related to the mass-spectra for Kaluza-Klein
compactifications of Supergravity on $AdS_5\times S^5$ and $AdS_7\times S^4$. Our approach
interprets the partition function of fields on AdS spacetime
in terms of wave-functionals that satisfy the 
functional \Sc equation.

\end{abstract}

\end{titlepage}

\section{Introduction}

Given the current interest in Supergravity in an Anti de Sitter background it
would seem useful to develop techniques for studying quantum field theory
beyond tree-level in such a spacetime. In this paper we will
consider the simplest one-loop quantity, namely the conformal anomaly
of the boundary theory. The central object of study in the AdS/CFT correspondence,
\cite{Maldacena} is a functional integral for a quantum field theory 
in Anti de Sitter space expressed in terms of the boundary values
of the field. Whilst this can be treated by the usual semiclassical
expansion it may also be interpreted as
the {\it large} time limit of the \Sc functional of the quantum field theory
and so satisfies a functional \Sc equation. We will show how to 
obtain this large time behaviour from a short time expansion using 
analyticity. We will illustrate our method by reproducing
the known two-point functions of the CFT for free scalar and fermionic fields 
in AdS and then
derive the one-loop contribution to the conformal anomalies for these theories.
Our approach also yields a straightforward calculation of the tree-level
contribution to the conformal anomaly for the pure gravity sector.

Following \cite{Witten} we consider the Euclidean version of $AdS_{d+1}$ with co-ordinates $\{ x^\mu \}\equiv \{t,x^1,..,x^d\}$ and metric
\be
ds^2={1\over (x^0)^2}\sum_{\mu=0}^d (dx^\mu)^2={1\over t^2}(dt^2+d\x \cdot d\x).
\label{met}
\ee
We will think of $t$, which is restricted to the range $t>0$, as Euclidean time. The boundary, $\partial M$, consists of $R^d$ at $t=0$ conformally compactified 
to a sphere by adding a point 
corresponding to $t=\infty$ where the metric vanishes. For illustration consider
a scalar field theory propagating on this space-time. We study the
functional integral
\be
Z[\varphi]=\int {\cal D}\phi\,e^{-S}\,{\Big |}_{\phi|_{\partial M}=\varphi},
\quad \quad S={\textstyle 1\over 2}\int d^{d+1}x\,{\sqrt g}\,\left( g^{\mu\nu}\partial_\mu\phi\,\partial_\nu
\phi +V(\phi)\right)\label{Z}
\ee
where ${\cal D}\phi$ is the volume element induced by the reparametrization invariant inner product on variations of $\phi$, $||\delta\phi||^2=
\int dt\,d\x \,\delta\phi^2/t^{d+1}$. We will need to regulate this by restricting $t$ to the range $\tau >t>\tau'$,
so define
\be
\Psi_{\tau,\tau'}[\tilde\varphi,\varphi]=\int {\cal D}\phi\,e^{-S}\,{\Big |}_{\phi(\tau)=\tilde\varphi,\phi(\tau')=\varphi}.\ee
Since the point corresponding to $t=\infty$ is part of $\partial M$ we
set $\tilde\phi=\lim_{|\x|\rightarrow\infty}\varphi (\x)$ as we take the limit
of large $\tau$ and small $\tau'$ to recover $Z$.
This is usually described as a partition function, but it may also be interpreted in
terms of the wave-functionals that represent
states in the \Sc representation. 
First change variables from $t,\phi$
to $\bar t=\ln t,\,\bar\phi=\phi/t^{d/2}$ so that the volume element
and kinetic term become the usual ones associated with the canonical 
quantization of $\bar\phi$. 
Thus
$Z[\varphi]$ is the $\tau '\rightarrow 0,\,\tau\rightarrow \infty$ limit
of
\be
\Psi_{\tau,\tau'}[\tilde\varphi,\varphi]=\int {\cal D}\bar\phi\,e^{-\bar S-S_b}\equiv \bar Z[\bar\phi_f,\bar\phi_i]\,e^{-S_b}
\ee
where
\be
\bar S={1\over 2}\int d\bar t\,d\x\left(\left({\partial\bar\phi\over\partial\bar t}\right)^2+
{d^2\over 4}\bar\phi^2 +t^2\nabla\phi
\cdot\nabla\phi +t^{-d}\,V(\bar\phi \,t^{d/2})\right),
\quad 
S_b={d\over 4}(\bar\phi_f^2-\bar\phi_i^2)
\ee
and $\bar\phi_f$, $\bar\phi_i$ are the value of $\bar \phi$ on the surfaces
$\bar t= \bar t_1=\ln \tau$ and $\bar t= \bar t_2=\ln \tau'$ respectively.
Now $\bar Z[\bar\phi_f,\bar\phi_i]$ can be interpreted as the \Sc functional,
i.e. the
matrix element of
the time evolution operator between eigenstates of the field operator,

\be
\bar Z[\bar\phi_f,\bar\phi_i]=\langle \,\bar\phi_f\,|
\, T\, \exp(-\int_{\bar t_2}^{\bar t_1} d t\, H ( t) )\,|\,\bar\phi_i\,\rangle\, ,
\label{fun}
\ee
which satisfies the functional \Sc equation
\be
{\partial\over \partial\bar t_1}\bar Z[\bar\phi_f,\bar\phi_i]
=-{\textstyle 1\over 2}\int d\x\,\left (-\,{\delta^2\over\delta\bar\phi_f^2} +t^2\, \nabla\bar\phi_f
\cdot\nabla\bar\phi_f + {d^2\over 4}\bar\phi_f^2 +t^{-d}\,V(\bar\phi_f \,t^{d/2})\right)\,\bar Z[\bar\phi_f,\bar\phi_i]
,\label{Schr''}
\ee
with the initial condition that it tends to a the delta-functional
$\delta[\bar\phi_f-\bar\phi_i]$ as $\bar t_1$ approaches $\bar t_2$.
We can re-write this in terms of the boundary values of our
original variables $t,\phi$, i.e. $\tau$ and $\tilde\varphi$
\be
{\partial\over \partial\tau}\Psi_{\tau,\tau'}[\tilde\varphi,\varphi]
=-{\textstyle 1\over 2}\int d\x\,\left (-\Omega^{-1}\,{\delta^2\over\delta\tilde\varphi^2} +\Omega\, \nabla\tilde\varphi
\cdot\nabla\tilde\varphi +\Omega' \,V(\tilde\varphi) +{\cal E}/\tau\right)\Psi_{\tau,\tau'}[\tilde\varphi,\varphi],\label{Schr}
\ee
where $\Omega=\tau^{1-d}$, $\Omega'=\Omega/\tau^2$.
${\cal E}$ arises from the action of the Laplacian on 
$S_b$,
formally

\be
{\cal E}=-\tau^d\,{\delta^2\over\delta\tilde\varphi^2}S_b
=-{d\over 2}\delta^d(0)
\ee
Clearly
the coincident functional derivatives stand in need of regularization,
which introduces a short-distance cut-off. By extending the flat-space 
arguments of Symanzik \cite{Symanzik} we would expect that for a renormalizable
field theory wave-function renormalization and an appropriate choice of the dependence of $V$ on the cut-off would ensure the finiteness
of the solution to (\ref{Schr}) in the limit that the cut-off is removed.
Since this can involve the use of counter-terms associated with the boundaries
we would expect that the renormalization constants may depend on $\tau$ and $\tau'$.
However in many applications the tree-level solution is sufficent for which
these considerations are unnecessary. We will see later that ${\cal E}$ contributes to the
conformal anomaly.
A similar argument yields the \Sc equation that gives the $\tau'$ dependence
\be
-{\partial\over \partial\tau'}\Psi_{\tau,\tau'}[\tilde\varphi,\varphi]
=-{\textstyle 1\over 2}\int d\x\,\left (-\Omega^{-1}\,{\delta^2\over\delta\varphi^2} +\Omega\, \nabla\varphi
\cdot\nabla\varphi +\Omega' \,V(\varphi) -{\cal E}/\tau\right)\Psi_{\tau,\tau'}[\tilde\varphi,\varphi],\label{Schr'}
\ee

In \cite{Paul}-\cite{Marcos} a general approach to solving the functional \Sc equation was developed. 
Consider the logarithm of $\Psi_{\tau,\tau'}[\tilde\varphi,\varphi]$,
$W_{\tau,\tau'}[\tilde\varphi,\varphi]$.
In perturbation theory this is a
sum of connected Feynman diagrams. 
In Quantum Mechanics we might try to solve the \Sc equation
by expanding in a suitable basis of functions. An analogous construction
is to expand  in terms of local functionals, i.e. a derivative expansion.
Now $W_{\tau,\tau'}[\tilde\varphi,\varphi]$ 
is non-local, but 
it will reduce to the integral of an infinite sum of local terms
when it is evaluated for fields that vary slowly on the scale
of $\tau$. Each term will depend on a finite power of the field and its derivatives at a 
single point, $\bf x$, on the quantization surfaces $t=\tau,\,t=\tau'$.
Although this expansion is appropriate  for slowly varying fields the  functional for arbitrarily varying fields can be reconstructed from it because the functional evaluated for scaled fields $\tilde\varphi(\x/\sqrt\rho)$
and $\varphi(\x/\sqrt\rho)$
can be analytically continued to the complex $\rho$-plane with the negative real axis removed.
Thus Cauchy's theorem will allow us to relate rapidly varying fields (small $\rho$)
to slowly varying ones (large $\rho$), and from this we can use the behaviour for small $\tau$  to obtain that for large
$\tau$, which is what is needed in the 
AdS/CFT correspondence.
Similar considerations enable the \Sc equation,
which because it has an ultra-violet cut off involves rapidly varying fields, to be turned into an equation acting directly on the local expansion,
determining the coefficients of that expansion from a set of {\it algebraic} equations. (In flat space these equations can be solved perturbatively where they yield the usual results for short-distance effects in scalar field theory
\cite{PMJ} and for the beta-function in Yang-Mills theory \cite{Marcos}.)

In the next section we prove the analyticity property that we will need to relate 
short and large time behaviour. We apply this to the example of the scalar field in section 3, reproducing the known scaling property of the two-point function in the 
CFT, and in section 4 we compute the one-loop conformal anomaly for this 
$d$-dimensional boundary theory, showing that it vanishes unless
$\sqrt{(d/2)^2+m^2}$ is  
a positive integer, where $m$ is the mass
in the AdS Lagrangian.
For $d=2$ this integer is the value of the central charge of the 
Virasoro algebra.
The free fermion is discussed in section 5 where we
show that the chiral projection that is known to occur in this theory is
related to the reducibility of the Floreanini-Jackiw representation of fermionic
fields. We also compute the conformal anomaly at one-loop for fermions
and find that similarly to the scalar case, it vanishes unless $|m|+1/2$ is an integer. For $d=2$ this integer
is again the value of the central charge of the Virasoro algebra.
For $d=4$ these values of the masses correspond to the scalar and fermion mass spectra in the Kaluza-Klein compactification of Supergravity on $AdS_5\times S^5$.
Finally, in section 6 we discuss pure gravity, for which our formalism yields a simple calculation of the
conformal anomaly.

\section{Analyticity of \Sc functional}

To show analyticity of $\Psi_{\tau,\tau'}[\tilde\varphi,\varphi]
$ we generalize to curved space the arguments of \cite{Paul}-\cite{Marcos}. We first make the dependence on $\tilde\varphi,\,\varphi$
explicit by modifying a standard phase-space derivation of the functional 
integral representation of the \Sc functional. By the usual argument we can write $Z[\bar\phi_f,\bar\phi_i]$
as

\be
{{\lim_{n\rightarrow \infty}}} \int \prod_{j=1}^n {\cal D}\bar\phi_j\,
\prod_{j=1}^{n+1} {\cal D}\Pi_j\,exp \left({\sum_{j=1}^{n+1}(-H [\Pi_j,\bar\phi_{j-1}]\,\delta\bar t_j)
+i\int d\x\,\Pi_j\,(\bar\phi_j-\bar\phi_{j-1})}\right)\,,
\label{phase}
\ee
where $\Pi$ is the eigenvalue of the canonical momentum conjugate to $\bar \phi$, $\delta \bar t_j=\bar t_j-\bar t_{j-1}$ and $\bar\phi_{n+1}=\bar\phi_f$
and $\bar\phi_0=\bar\phi_i$. $\bar\phi_f$ appears in this expression only in the
$i\Pi_{n+1}\bar\phi_{n+1}$ term in the exponent, whereas $\bar\phi_i$ appears
in $-i\Pi_{1}\bar\phi_{0}$, and terms proportional to $\delta\bar t_1$, which can be neglected as $\delta\bar t_1\rightarrow 0$. So the contribution of
$\bar\phi_i$ and $\bar\phi_f$ to (\ref{phase}) can be manifested
by adding to the exponent $i\int d\x\,(\Pi_{n+1}\,\bar\phi_f
-\Pi_1\,\bar\phi_i)$ and taking $\bar\phi_{n+1}=\bar\phi_0=0$.  Thus $\bar\phi_i$ and $\bar\phi_f$ appear as sources coupled to $\dot{\bar\phi}$ when we integrate out the momenta, $\Pi_j$, so that we arrive at the functional integral

\be
\int {\cal D}\bar\phi\,\exp\left( -\bar S+\int d\x \,\left(
\bar\phi_f \,\dot{\bar\phi} (\bar t_1)-\bar\phi_i\dot{\bar\phi} (\bar t_2)\right)\right)\,\exp\left(\int d\x \,\Lambda\,(\bar\phi_i^2+\bar\phi_f^2)\right)\,,\label{int}
\ee
where the boundary condition on $\bar\phi$ is now that it should vanish 
at $\bar t=\bar t_1,\,\bar t_2$. $\Lambda$ is a regularization of $1/\epsilon$ which cancels
certain divergences that arise in the evaluation of (\ref{int}) whose origin is explained in \cite{Paul}. If we now interchange the r\^oles of $\bar t$ and $x^1$, and think of
$x^1$ as Euclidean time and $\bar t, x^2,..x^d$ as spatial co-ordinates then 
we can give an alternative interpretation of the functional integral
(ignoring the $\Lambda $ factor for the time being) as
the vacuum expectation value 

\be
\langle \, O_r \, |T\, \exp (\int dx^1\,\varphi_i\hat R_i(x^1))\,|\,O_r\,\rangle\,,
\label{vev}
\ee
where $|\,O_r\,\rangle$ is the vacuum for the Hamiltonian, $\hat H_r$, associated with the
quantization surfaces of constant $x^1$ and $\bar t_2< \bar t<\bar t_1$. Unlike
the previous Hamiltonian which depended on $\bar t$, this operator is independent of
its associated `time', $x^1$. $\varphi_i$ is a compact notation for the sources,
so that

\be
\varphi_i\hat R_i(x^1)\equiv
\int dx^2..dx^d\,\left(
\bar\phi_f \,\dot{\bar\phi} (\bar t_1)-\bar\phi_i\dot{\bar\phi} (\bar t_2)\right)
\ee
Expanding (\ref{vev}) in powers of the sources and using $\hat H_r$ to generate the
$x^1$-dependence of the operators $\hat R_i$ gives

\bq
&&\langle \, O_r \, |T\, \exp \left(\int dx^1\,\varphi_i
\hat R_i(x^1)\right)\,|\,O_r\,\rangle=\nonumber\\
&&
\sum_{n=0}^\infty
\int_{-\infty}^\infty dx^1_n\int_{-\infty}^{x^1_n}dx^1_{n-1}..
\int_{-\infty}^{x^1_3} dx^1_2\int_{-\infty}^{x^2_2}dx_1^1
\,\prod_{j=1}^n \varphi_{i_j}(x^1_j)\nonumber\\
&&
\times\langle \, O_r \, |\,\hat R_{i_n}(0)\,e^{(x_{n-1}^1-x_n^1)\,\hat H_r}\,\hat R_{i_{n-1}}(0)\, 
..\,\hat R_{i_2}(0)\,e^{(x_{2}^1-x_1^1)\,\hat H_r}\,\hat R_{i_1}(0)\,|\,O_r\,\rangle
\label{expll}
\eq
We have taken the eigen-value of $\hat H_r$ belonging to
$|\,O_r\,\rangle$ to be zero.
Fourier transforming the $x^1$-dependence of the sources as
$\varphi_i(x^1)=\int dk \tilde\varphi_i (k)\exp({-ikx^1})$
enables the $x^1$ integrals to be done yielding

\bq
&&
\sum_{n=0}^\infty
\,\delta(\,\sum_{j=1}^nk_j\,)\,\prod_{j=1}^n \int dk_j\,\tilde\varphi_{i_j}(k_j)\nonumber\\
&&
\times\langle \, O_r \, |\,\hat R_{i_n}(0)\,{1\over \hat H_r-i\sum_1^{n-1}k_j} 
\,\hat R_{i_{n-1}}(0)\, ..\,\hat R_{i_2}(0)\,{1\over \hat H_r-ik_1}\,\hat R_{i_1}(0)\,|\,O_r\,\rangle
\label{expl}
\eq
Suppose that we had computed the \Sc functional for new sources obtained by scaling $x^1$, $\varphi_i(x^1/\sqrt \rho,x^2,..x^d)$,
with $\rho$ real and positive. Then we would have obtained the same 
expression as (\ref{expl}) but multiplied by $\sqrt\rho$ and with the 
$\hat H_r$ in the denominators replaced
by $\sqrt\rho\hat H_r$. We took $\rho$ to be real and positive,
but we can use this expression to continue to the complex $\rho$-plane. 
Since the eigenvalues of $\hat H_r$ are real we conclude that
the result is analytic in the whole plane with the negative real axis removed.
This assumes that we work to finite order in the sources, and that
the spectral decomposition of (\ref{expl}) as a sum over eigen-values of
$\hat H_r$ converges, as we should expect if the \Sc functional is finite.
The terms in $\Lambda$ in (\ref{int}) do not affect this conclusion.
By repeating the argument with $x^1$ interchanged with each of the other
co-ordinates in turn we conclude that 
for $\varphi^\rho_i(\x)\equiv \varphi_i(\x/\sqrt\rho)$
the \Sc functional $Z[\bar\phi_f^\rho,\bar\phi_i^\rho]$ and consequently $\Psi_{\tau,\tau'}[\tilde\varphi^\rho,\varphi^\rho]$
are (to any finite order in the sources) analytic in $\rho$
in the plane cut along the negative real axis. Since
$\x\rightarrow\x/\sqrt\rho,\,t\rightarrow t/\sqrt\rho$ is
an isometry of (\ref{met}) it follows that 

\be\Psi_{\tau,\tau'}[\tilde\varphi^\rho,\varphi^\rho]=
\Psi_{\tau/\sqrt\rho,\,\tau'/\sqrt\rho}[\tilde\varphi,\varphi]
\label{iso}\ee
and we see that analyticity in $\rho$ corresponds to the analyticity in
time associated with Wick rotation. Subject to the caveat that we work to 
finite order in the sources the logarithm $\Psi_{\tau,\tau'}[\tilde\varphi^\rho,\varphi^\rho]$, $W_{\tau,\tau'}[\tilde\varphi,\varphi]$, is just a sum of products of terms appearing in (\ref{expl}) so it too is analytic in the cut
$\rho$-plane when it is evaluated for the scaled sources. 
 
We can use (\ref{expl}) to justify a local expansion for 
$W_{\tau,\tau'}[\tilde\varphi,\varphi]$ with a non-zero radius of convergence. Since $\hat H_r$ is the Hamiltonian 
for a field theory on a finite `spatial' interval such that $\hat \phi$
vanishes at the ends there is a mass-gap of the order of $1/\tau$ for small
$\tau$. When the momenta $k_j$ are sufficently small
on the scale of this mass-gap
we can expand the denominators in the spectral decomposition of 
(\ref{expl}) in integer powers of $\sum k$,
except for the contribution of the vacuum,
which is of the form $1/\sum k$. The singular behaviour as $k\rightarrow 0$
must disappear when we take the logarithm to ensure
cluster decomposition. We conclude that any term of finite order in the
sources in $W_{\tau,\tau'}[\tilde\varphi,\varphi]$ has a local 
expansion for $\tilde\varphi,\varphi$ that vary sufficently slowly 
with $\x $. This will take the form
$W_{\tau,\tau'}[\tilde\varphi,\varphi]=\int d^d\x(a\varphi^2 +b\tilde\varphi^2
+c\varphi\nabla^2\varphi+d\varphi\nabla^2\nabla^2\varphi..)$ with $a,\,b,\,c,..$ depending on $\tau,\,\tau'$.
If the Fourier transforms of the sources have bounded support then
$\tilde\varphi^\rho,\varphi^\rho$ are slowly varying for large $\rho$.
Since each term in the expansion of $W_{\tau,\tau'}[\tilde\varphi^\rho,\varphi^\rho]$ has a simple dependence
on $\rho$, we have that for large $\rho$ the functional $W_{\tau,\tau'}[\tilde\varphi^\rho,\varphi^\rho]={\sqrt\rho}^d\int d^d\x\,(a\varphi^2 +b\tilde\varphi^2
+c\varphi\nabla^2\varphi/\rho+d\varphi\nabla^2\nabla^2\varphi/\rho^2..$.
The powers of $\rho$ are $d/2+{\rm integer}$ so we conclude that
the cut on the negative real axis in
${\sqrt\rho}^dW_{\tau,\tau'}[\tilde\varphi^\rho,\varphi^\rho]$
runs only a finite distance from the origin.

We will now exploit the analyticity to obtain the logarithm of the partition function $\log Z[\varphi]=\lim_{\tau\rightarrow\infty,\tau'\rightarrow 0}W_{\tau,\tau'}[0,\varphi]$ from the small $\tau$ behaviour, which
is computable from a power series solution of the \Sc equation.
(We take $\tilde\p=\lim_{\x\rightarrow\infty}\p=0$.)
For simplicity we assume continuity in $\tau'$ at the origin, and consider
$W_{\tau/\sqrt\rho,0}[\tilde\varphi,\varphi]$. In general there will need to be
some $\tau$-dependent renormalization in order that the limit as $\tau\rightarrow \infty$ exists, including wave-function renormalization,
$\varphi_{\rm ren}=\sqrt{z(\tau)}\p$, this will be the case even for free `massive' fields at tree-level. Given that $\p$ is a scalar the isometry
$x^\mu\rightarrow\lambda x^\mu$ implies that the functional takes the form

\be
\int d^d\x\,{\Omega\over\tau}\left( a+{1\over 2}\tilde\varphi\,\Gamma (-\tau^2\nabla^2)\,\tilde\varphi+\tilde\varphi\,\Xi (-\tau^2\nabla^2)\,
\varphi+{1\over 2}\varphi\,\Upsilon (-\tau^2\nabla^2)\,\varphi+..\right)
\label{expand}
\ee
where the dots stand for terms of higher order in the fields of which the general term is of the form
\be
{\Omega\over\tau}\int d^d\x\,\Gamma_{m,n}(\tau\nabla_1,..,\tau\nabla_{n+m})
\varphi(\x_1)..\varphi(\x_n)\,\tilde\varphi(\x_{n+1})..\tilde\varphi(\x_{n+m})
{\Big|}_{\{\x_i=\x\}}
\label{gen}
\ee
At short times, or equivalently, for slowly varying fields, we have the local
expansions
\be
\Gamma =\sum_{n=0}^\infty b_n\,(-\tau^2\nabla^2)^n,\quad
\Xi =\sum_{n=0}^\infty c_n\,(-\tau^2\nabla^2)^n,\quad
\Upsilon =\sum_{n=0}^\infty f_n\,(-\tau^2\nabla^2)^n,
\label{local}
\ee
with $b_n,\,c_n,\,f_n$ constants. Renormalizability would imply that
\be
{1\over \tau^d z(\tau)}\left[ \Upsilon (-\tau^2 \nabla^2)+{\rm polynomial\,\, in}\,\,\nabla
\right]\label{twopoint}
\ee
is finite as $\tau\rightarrow\infty$. Suppose that for large $\tau$ the renormalization constant depends on $\tau$ as $z(\tau) \sim\tau^{2q}$
then finiteness of the limit of (\ref{twopoint}) requires that
for large $\tau$, $\Upsilon(-\tau^2 \nabla^2)\sim (-\tau^2\,\nabla^2)^{d/2+q}\upsilon$,
and our problem is to calculate $\upsilon$ and $q$.
The general term in (\ref{gen}) 
should depend on $\tau$ as $\Gamma_{m,n}(\tau\nabla_1,..,\tau\nabla_{n+m})\sim
\tau^{d+(n+m)q} F_{m,n}(\nabla_1,..,\nabla_{n+m})$  and we need to calculate $F_{m,n}$. Now
our previous arguments imply that $\Upsilon(1/\rho)$ is
analytic in the complex plane with a finite cut extending from the origin along the negative real axis so we can evaluate the following integral

\be
I(\lambda)={1\over 2\pi i}\int_C {d\rho\over \rho}\,e^{\lambda \rho}
\,  \Upsilon(1/\rho)\label{I}
\ee
in two ways. We take $C$ to be a circle centred on the origin and large enough for us to
be able to use the local expansions (\ref{local}) to give

\be
I(\lambda)=\sum_{n=0}^\infty {f_n\lambda^n\over n!}
\ee
The integral may also be evaluated by collapsing the contour
$C$ onto the cut. Let this consist of a small circle about the origin,
of radius $\eta$,
and two lines close to the negative real axis running from the circle
to the end of the cut. The contribution from the latter is suppressed
if the real part of $\lambda$ is large and positive.
That from the circle is controlled by
the large time behaviour

\be
{1\over 2\pi i}\int_{|\rho|=\eta} {d\rho}\,{\upsilon\,e^{\lambda \rho}
\over \rho^{d/2+q+1}}=\upsilon\lambda^{d/2+q}\left({1\over 2\pi i}\int_{|\rho|=|\lambda|\eta} {d\rho}\,{e^{\rho}\over \rho^{d/2+q+1}}\right)
\ee
and for large $\lambda$ this is ${\lambda^{d/2+q}/\Gamma(d+2q+1)}$.
So, as the real part of $\lambda$ tends to $+\infty$ we obtain

\be
I(\lambda)=\sum_{n=0}^\infty {f_n \lambda^n\over n!}\sim {\upsilon\lambda^{d/2+q}\over (d+2q)!}
\label{summo}
\ee
enabling us to compute $\upsilon$ and $q$ from a knowledge of the local expansion
alone. Now for positive real $\lambda$, $\sum_{n=0}^\infty {f_n \lambda^n}$ is an alternating series with finite radius of convergence. By comparing terms with
those of $\exp (-{\rm constant} \lambda)$ it follows that this converges
for all $\lambda$ and so we can take $\lambda$ large, even though this series
is in positive powers of $\lambda$. Furthermore, as we shall see in some examples below,  we obtain a good approximation to the large $\lambda$ limit by truncating the series at some order, and then taking $\lambda$ as large as is consistent with the truncation, i.e. so that the term of highest order in
$\lambda$ is a small fraction of the sum.
This generalizes in an obvious way to the other terms in (\ref{expand}).

\vfill
\eject

\section{Example: free scalar field}

As an example consider the free massless theory so $V=0$ in (\ref{Z});
 the action is

\be S={1\over 2}\int d^d\x\,dt\,\left(\Omega\sum_{\mu=0}^d(\partial_\mu\phi)^2
\right)
\label{free1}
\ee
with $\Omega=1/t^{d-1}$. 
For this Gaussian functional integral only those terms shown explicitly in
(\ref{expand}) are present. Substituting this into the \Sc equation 
(\ref{Schr}) yields

\bq
&&
{\partial\over\partial\tau}\left({\Omega\over\tau}\Gamma\right)=
{\Omega\over\tau^2}\Gamma^2+\Omega\nabla^2,\quad\quad
{\partial\over\partial\tau}\left({\Omega\over\tau}\Xi\right)=
{\Omega\over\tau^2}\Gamma\Xi
\nonumber\\
&&
{\partial\over\partial\tau}\left({\Omega\over\tau}\Upsilon\right)={\Omega\over\tau^2}\Xi^2,\quad \quad{\partial\over\partial\tau}\left({\Omega\over\tau}a\right)=
{1\over 2\tau} \,\left(\Gamma +{{d}\over 2}\right)\delta^d(\x-\y)\Big|_{\x=\y}
\label{eqc}
\eq
These, together with the initial condition, lead to the recursive solution
of the coefficients of the local expansions (\ref{local})
\bq
&& b_0=-d=-c_0=f_0,\quad b_1=-1/(2+d)\nonumber\\
&& b_n={\sum_{q=1}^{n-1} b_q\,b_{n-q}\over 2n+d},\quad
c_n={ \sum_{q=1}^n b_q\,c_{n-q} \over 2n},\quad f_n={ \sum_{q=0}^{n} c_q\,c_{n-q} \over 2n-d}.
\label{localsol}
\eq
If we 
were to take $d$ to be an even positive integer the relations for the
$f_n$ would break down, thus keeping $d$ variable regulates
the solution for $f_n$. Note that  there is no need to solve these relations further as they are ideally suited to numerical evaluation of the coefficients.
\begin{figure}[h]
\unitlength1cm  
\begin{picture}(14,10)
\put(9,0.25){$\lambda$}
\put(12,5.5){$\tilde S_{50}$}
\put(12,3){$\tilde S_{49}$}

\centerline{\epsfysize=12cm
\epsffile{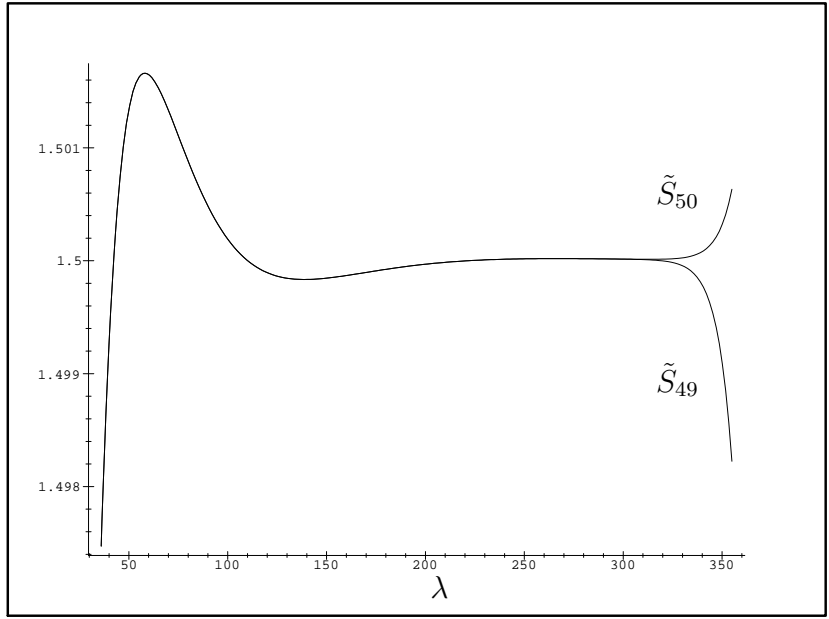} }
\end{picture}

\caption[wave1]{\label{wave1} The truncated series as a function of $\lambda$.}
\end{figure}

\begin{figure}[h]
\unitlength1cm  
\begin{picture}(14,10)
\put(9,1){$d$}
\put(8,6){$q \times d^3$}
\centerline{\epsfysize=12cm
\epsffile{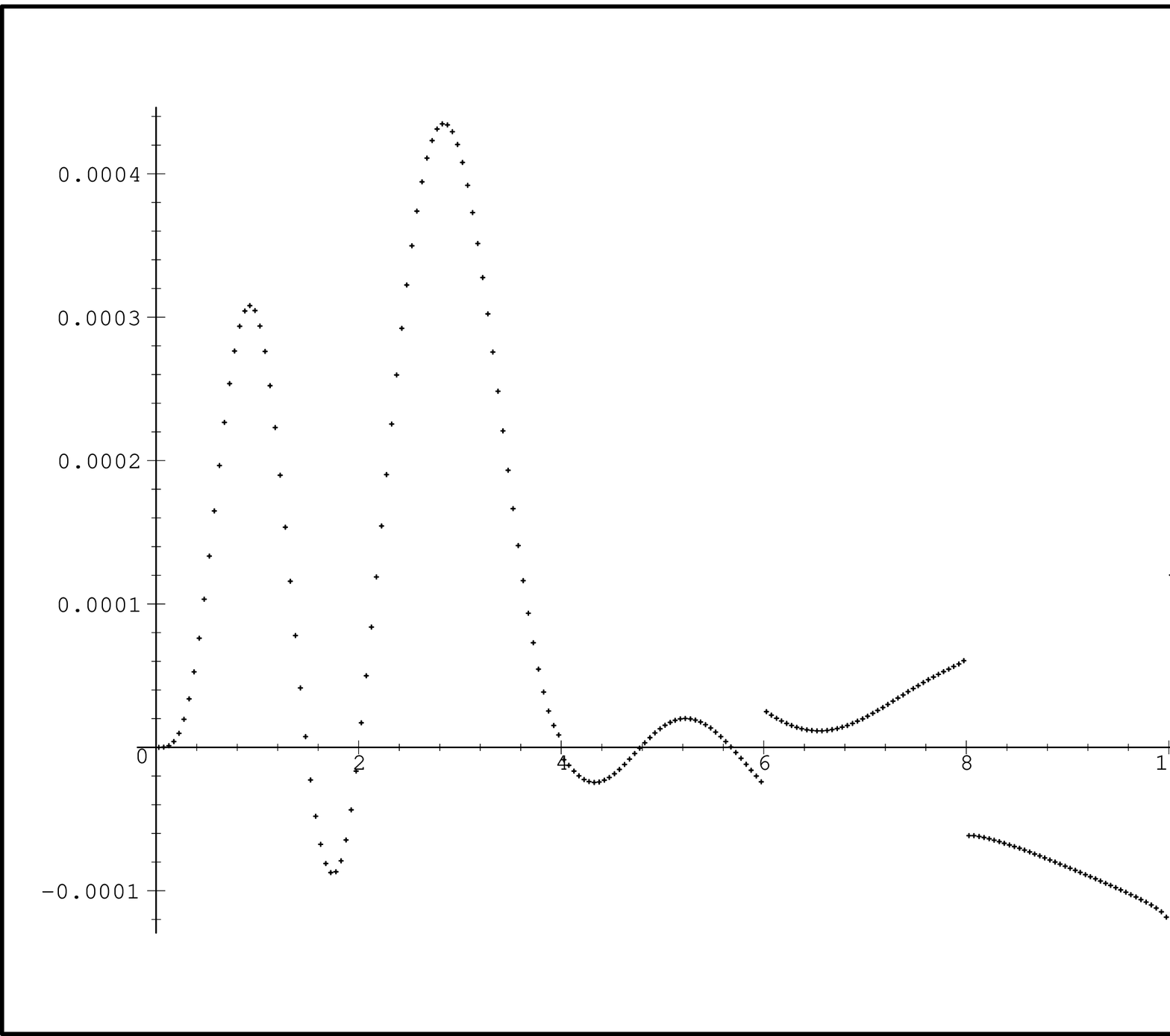} }
\end{picture}

\caption[wave2]{\label{wave2} $q$ as a function of $d$.}
\end{figure}

To illustrate the calculation of $\upsilon$ and $q$ take the example of $d=3$. 
Our discussion will involve numerical computations. Although the rest of this paper is concerned with analytic results the numerical work of this section
highlights the use of the derivative expansions and is readily
generalisable to interacting theories \cite{PMJ}.
From (\ref{summo}) we have that for large $\lambda$, $\tilde I(\lambda)\equiv d(\log (I(\lambda))/d(\log (\lambda))\rightarrow \tilde d/2+q$. Truncating the infinite series
to its first $N$ terms, $S_N(\lambda)$, gives an approximation to this. In Figure 1 we have shown $\tilde S_N\equiv d(\log (S_N(\lambda))/d(\log (\lambda))$
for $N=50$ and $N=49$. The two curves rapidly settle down to
a value of approximately $1.5$ for $\lambda>50$ but separate noticeably at $\lambda$
$\approx 300$ above which the truncated series cease to be good approximations to $I(\lambda)$. We estimate $d/2+q$ by taking $\lambda=290$ where the separation
between the two curves is about $0.5\times 10^{-6}$, which is much less
than the error ${\cal E}$ obtained by approximating the limiting value of $\tilde I$
by its value for finite $\lambda$. This gives  $d/2+q\approx \tilde S_{50}
(290)=1.500017$. The error is obtained by
studying how $I(\lambda)/\lambda^{d/2+q}$ settles down to a constant value. 
For small $\lambda$ the approach to a constant value is controlled by exponential terms that originate from the suppression of the contribution of the cut, but for larger $\lambda$ the error is dominated by power corrections to the small $\rho$ behaviour 
of $\Upsilon (1/\rho)$. A plot of $(S_{50}(\lambda)/\lambda^{1.5}-S_{50}(290)/290^{1.5})\lambda^{2.6}$
reveals oscillations of roughly equal amplitude approximately equal to $12$,
so that the error in approximating the $\lambda\rightarrow\infty$ value of $I(\lambda)/\lambda^{d/2+q}$ is of order $12/\lambda^{2.6}$ leading to an estimate of the error ${\cal E}=\pm2\times 10^{-5}$. So we conclude that 
$q=0$ to the accuracy of our calculations. Having obtained $q$ we can estimate
$\upsilon$ from the $\lambda=290$ value of $S_{50}\Gamma(2.5)/\lambda^{1.5}$
as $0.999998$ with an error of $\pm 10^{-5}$. 
We have repeated the calculation of $q$ for various values of 
$d$.  The results are shown in Figure 2. We have plotted our estimate of
$q$ (multiplied by $d^3$ to make the results for large $d$ visible)
for $d$ varying in steps of $0.025$ from $0.05$ to $10.5$.
The results are consistent with $q=0$, which is the exact result 
of \cite{Witten}. By calculating a few values of $\upsilon$ we 
guessed that its dependence on $d$ is given by

\be
\upsilon={-((d-3)/2)!^2 \,2^{d-3}\over  \sin (d\pi/2)\, (d-2)!^2}\equiv
\tilde\upsilon (d),
\label{upsilo}
\ee
and we test this by plotting in Fig 3 our numerical estimates of $\upsilon$
divided by the right hand side of (\ref{upsilo}) for the same range
of $d$ as before. From this, and $q=0$ we conclude that the 
$\varphi$-dependence of the AdS partition function is given by 

\be
\log\,Z[\varphi]=\lim_{\tau\rightarrow\infty} {1\over 2\tau^d}\int
d^d\x \,\varphi\, \tilde\upsilon(d)\,(-\tau^2\,\nabla^2)^{d/2}\,\varphi
\ee
This is a {\it non-local} expression, even when $d$ is an even integer, 
thanks to the singularity that then appears in $\tilde\upsilon$,
since

\be
\tilde\upsilon(d)\,(-\nabla^2)^{d/2}\,\delta^d(\x-\y)={\tilde c\over |\x-\y|^{2d}},
\quad \tilde c={(d-1)\,(d/2)!\,(d/2-3/2)!^2\over 8\,
\pi^{d/2+1}\,(d-2)!}
\ee
so that
\be
\log\,Z[\varphi]= {\tilde c\over 2}\int
d^d\x \,d^d\y\,{\varphi (\x)\, \varphi(\y)\over |\x-\y|^{2d}}
\ee
as in \cite{Witten}. 

\begin{figure}[h]
\unitlength1cm  
\begin{picture}(14,10)
\put(9,0.25){$d$}
\put(9,4.5){$\upsilon/\tilde\upsilon$}

\centerline{\epsfysize=12cm
\epsffile{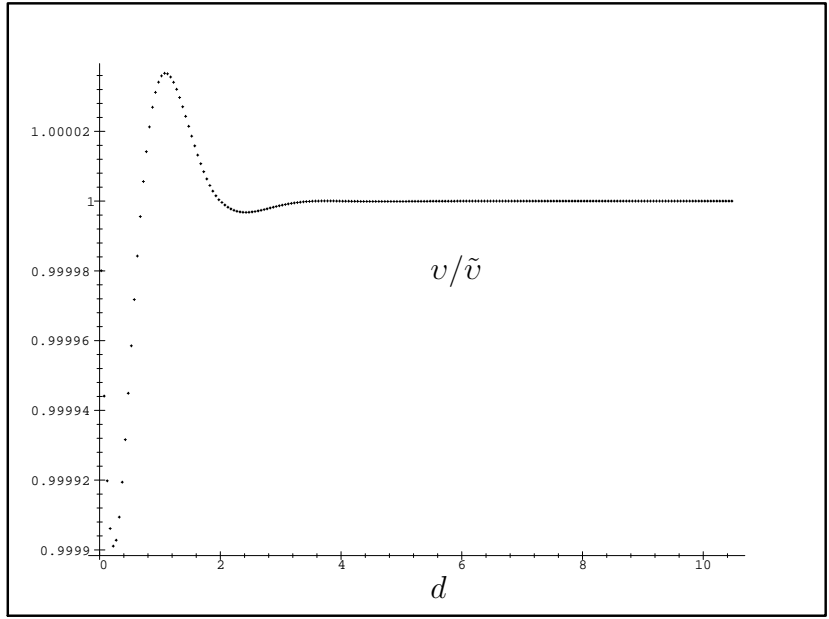} }
\end{picture}

\caption[wave3]{\label{wave3} $\upsilon$ as a function of $d$.}
\end{figure}

The calculation of $a$ in (\ref{eqc}) requires the introduction of
a regulator into the functional Laplacian of (\ref{Schr}). We
will do this with a cut-off on the eigenvalues, $k^2$ of $-\nabla^2$,
restricting them to be less than $1/(\tau^2 s)$ where $s$ is the square of a fixed proper distance. Thus we replace $\delta^d(\x)$ in (\ref{eqc}) by

\be 
\theta\left(s\tau^2\nabla^2+1\right)\,\delta^d(\x)=\int_{k^2<1/(\tau^2 s)}{d^dk\over (2\pi)^d}e^{i{\bf k}\cdot \x}
\ee
where $\theta$ is the step-function,
giving

\be
a(s)=-{\tau^d\over 2d}\int_{k^2<1/(\tau^2 s)} d^dk\, \left(\Gamma(\tau^2k^2)+{d\over 2}\right)
=-{1\over 2d}\int_{k^2<1/s} d^dk\, \left(\Gamma(k^2)+{d\over 2}\right)
.
\ee
The continuum limit corresponds to taking $s$ to zero. Unfortunately
when we substitute our local expansion (\ref{local}) and (\ref{localsol})
into this we obtain a series that will converge only for large
values of $s$

\be
a(s)=-{V_d\over 2ds^{d/2}}\sum_{n=0}^\infty {b_n\over s^n \,(2n+d)}
\ee
where $V_d/d$ is the volume of the unit ball in $d$-dimensions.
However our previous arguments imply that $a(s)$ is an analytic function
of $s$ in the complex $s$-plane cut along the negative real axis, so that if
$a(s)\sim a_0/s^\nu$ for small $s$ then for large $\lambda$
\be
{1\over 2\pi i}\int {ds\over s}\,e^{\lambda s}\,a(s)=
-{V_d\over 2d}\sum_{n=0}^\infty {b_n\,\lambda^{n+d/2}\over  \,(2n+d)\,(n+d/2)!}
\sim {a_0\,\lambda^\nu\over \nu!}
\ee
Numerical investigation of this suggests that $\nu=d+1$.
The ultraviolet divergence of $a$ can be cancelled by renormalizing the cosmological constant.

The effect of adding a mass term to the action, $V(\phi)=m^2\,\phi^2$
can be understood quite simply by a change of variables back to the massless
action. If we set $\phi=t^{-r}\psi$ in the action

\be
S={1\over 2}\int_0^\tau dt\int d^d\x\,\left(t^{1-d}\sum_{\mu=0}^d\,(\partial_\mu\phi)^2
+t^{-d-1}\,m^2\,\phi^2\right),
\ee
it becomes
\bq
S
=&&{1\over 2}\int_0^\tau dt\int d^d\x\,\left(t^{1-d-2r}\sum_{\mu=0}^d(\partial_\mu\psi)^2
+t^{-d-1-2r}\,(m^2-r^2-rd)\,\psi^2\right)\nonumber\\
&&
-{1\over 2}\int d^d\x\,
r\tau^{-d-2r}\psi^2(\tau,\x)\label{massc}
\eq
so that if we take $r(r+d)=m^2$ we are left with a boundary term
plus the massless action (\ref{free1}) with $\Omega=1/t^{d+2r-1}\equiv 1/t^p$,
(The inner product on variations of $\psi$, from which we can construct the 
functional integral volume element ${\cal D}\psi$, is $||\delta\psi||^2=
\int dt\,d\x \,\delta\psi^2/t^{p+2}$.)
Consequently if we express $W_{\tau,0}[\tilde\varphi,\varphi]$
in terms of $\psi$ it takes the form corresponding 
to (\ref{expand})

\be
\int d^d\x\,{\Omega\over\tau}\left( a+{1\over 2}\tilde\psi\,\left(\Gamma (-\tau^2\nabla^2)-r\right)\,\tilde\psi+\tilde\psi\,\Xi (-\tau^2\nabla^2)\,
\psi+{1\over 2}\psi\,\Upsilon (-\tau^2\nabla^2)\,\psi\right)
\label{expandpsi}
\ee
where, as before $\Gamma,\,\Xi$ and $\Upsilon$ have the local/short-time expansions (\ref{local}) with solutions (\ref{localsol})
leading to 
\be
\log\,Z[\varphi]=\lim_{\tau\rightarrow\infty} {1\over 2\tau^{p+1}}\int
d^d\x \,\psi\, \tilde\upsilon(p+1)\,\left((-\tau^2\,\nabla^2)^{(p+1)/2}-r\right)\psi
\ee
A non-trivial limit occurs for positive $p+1$, which means that
$r$ is the positive root of $r(r+d)=m^2$, giving
\be
\log\,Z[\varphi]= {c'\over 2}\int
d^d\x \,d^d\y\,{\varphi (\x)\, \varphi(\y)\over |\x-\y|^{2d+2r}}
\ee
since $d+p+1=2(d+r)$, agreeing with \cite{Witten}. Taking this limit required wave-function renormalization 
of $\p$ with $z(\tau)=\tau^r$.

\medskip

 Having obtained $\Psi_{\tau,0}[\tilde\varphi,\varphi]$ we can find $\Psi_{\tau,\tau'}[\tilde\varphi,\varphi]$ using the self-reproducing property of the \Sc functional,

\be
\Psi_{\tau,0}[\tilde\varphi,\varphi]=
\int {\cal D}\hat\varphi\,\Psi_{\tau,\tau'}[\tilde\varphi,\hat\varphi]\,\Psi_{\tau',0}[\hat\varphi,\varphi].
\ee
If we denote the logarithm of $\Psi_{\tau,\tau'}[\tilde\varphi,\varphi]$ by

\be
\int d^d\x\,\left( a_{\tau,\tau'}+{1\over 2}\tilde\varphi\,\Gamma_{\tau,\tau'}\,\tilde\varphi+\tilde\varphi\,\Xi_{\tau,\tau'}\,
\varphi+{1\over 2}\varphi\,\Upsilon_{\tau,\tau'}\,\varphi\right)
\label{expand'}
\ee
then computing the Gaussian integral leads to 
\bq
&&
\Upsilon_{\tau,0}=\Upsilon_{\tau',0}
-(\Upsilon_{\tau,\tau'}+\Gamma_{\tau',0})^{-1}\Xi_{\tau',0}^2\nonumber\\
&&
\Xi_{\tau,0}=-\Xi_{\tau',0}\,(\Upsilon_{\tau,\tau'}+\Gamma_{\tau',0})^{-1}\Xi_{\tau,\tau'}
\nonumber\\
&&
\Gamma_{\tau,0}=\Gamma_{\tau,\tau'}
-(\Upsilon_{\tau,\tau'}+\Gamma_{\tau',0})^{-1}\Xi_{\tau,\tau'}^2\eq
hence
\bq 
&&
\Upsilon_{\tau,\tau'}=(\Upsilon_{\tau',0} -\Upsilon_{\tau,0})^{-1} \,\Xi_{\tau',0}^2-\Gamma_{\tau',0}
\nonumber\\&&
\Xi_{\tau,\tau'}=-(\Upsilon_{\tau',0} -\Upsilon_{\tau,0})^{-1} \,\Xi_{\tau,0}\,\Xi_{\tau',0}\nonumber\\&&
\Gamma_{\tau,\tau'}=\Gamma_{\tau,0}+(\Upsilon_{\tau',0} -\Upsilon_{\tau,0})^{-1} \,\Xi_{\tau,0}^2
\label{bigger'}
\eq
So, in terms of $\Gamma,\,\Xi$ and $\Upsilon$

\bq 
&&
\Upsilon_{\tau,\tau'}={\tau'^{-(p+1)}}\left((\Upsilon(-\tau'^2\nabla^2)-({\tau'/\tau})^{p+1}\Upsilon(-\tau^2\nabla^2))^{-1} \,\Xi(-\tau'^2\nabla^2)^2-\Gamma(-\tau'^2\nabla^2)\right)
\nonumber\\&&
\Xi_{\tau,\tau'}=(\tau\tau')^{-(p+1)}\left(\tau^{-(p+1)}\Upsilon(-\tau^2\nabla^2)
-\tau'^{-(p+1)}\Upsilon(-\tau'^2\nabla^2)\right)^{-1} \,\Xi(-\tau^2\nabla^2)\,\Xi(-\tau'^2\nabla^2)\nonumber\\&&
\Gamma_{\tau,\tau'}=\tau^{-(p+1)}\left(\Gamma(-\tau^2\nabla^2)+((\tau/\tau')^{p+1}\Upsilon(-\tau'^2\nabla^2) -\Upsilon(-\tau^2\nabla^2))^{-1} \,\Xi(-\tau^2\nabla^2)^2\right)
\label{bigger}
\eq

\vfill
\eject


\section{Conformal Anomaly for Scalar Fields}

So far we have worked
with a boundary sphere in which the curvature is effectively 
placed at spatial infinity. To discuss the conformal anomaly 
it will be more convenient to 
smooth this out, so instead of the metric (\ref{met}) we will now
use

\be
ds^2={1\over t^2}\left( dt^2+\sum_{i,j}g_{ij}\,dx^i\,dx^j\right).
\label{newmet}
\ee
with $i,j=1..d$. We take $g_{ij}$ to be the metric of a $d$-dimensional sphere
of large radius, $r$, obtained, for example, by stereographic projection onto the plane parametrized by $\x$. The conformal anomaly measures the response
of the free energy, which is the field independent part of $\log Z$,
to a Weyl transformation of $g_{ij}$. We wish to compute this as $r\rightarrow\infty$
when we recover the AdS metric (\ref{met}). 
When the boundary is a two-dimensional sphere the free energy should change by  
$\int d^2\x\,{\sqrt g}R\,c\delta\rho /(48\pi)=c\delta\rho/6$ when $\delta g_{ij}=g_{ij}\,\delta\rho$
where $R$ is the curvature of the boundary and $c$ the central charge of the Virasoro algebra. More generally there will be a conformal anomaly when the boundary has an even number of dimensions, $2N$ say. We will continue to
keep $d$ a continuous variable allowing it to tend to $2N$ at the end of 
our calculations.
Now $\log Z[\p]=\lim_{\tau\rightarrow\infty,\tau'\rightarrow 0} W_{\tau,\tau'}[0,\p]$
so we need to compute $W_{\tau,\tau'}[0,\p]$ in the presence of the curved metric $g_{ij}$, which we can still do using our previous technique. Since $r\rightarrow\infty$
it will be sufficent to 
find the free energy from a derivative expansion.
In this section we consider the scalar field.
We will discuss a free massive theory,
because at one-loop the calculation is the same as for an interacting scalar theory. 
The \Sc equation takes the same form as before,
(\ref{Schr}),
provided that $\Omega$ and $\Omega'$ acquire a factor of $\sqrt{\det g}$
and that $\nabla$ is the covariant derivative constructed from $g_{ij}$,
and the solution is again of the form (\ref{expand}), but with $a$ no
longer constant, but depending on $g_{ij}$ and $\tau$.
If we set $g_{ij}=\delta_{ij}+h_{ij}(\x)$, and treat $h_{ij}$ as
a source in the same way that we treated $\p$ and $\tilde\p$ as sources,
we can generalize our earlier discussion to argue that
$W_{\tau,\tau'}[\tilde\p^\rho,\p^\rho,g_{ij}^\rho]$ is analytic in
the cut $\rho$-plane, where $g_{ij}^\rho (\x)=g_{ij}(\x/\sqrt\rho)$.
Again, this allows us
to reconstruct the large $\tau$ solution of the \Sc equation from the small $\tau$ solution for which we have the local expansion 
(\ref{local}). 
By using $g_{ij}$ the \Sc functional can be made invariant under
reparametrizations of the space-like variables, giving

\be
W_{\tau,\tau'}[\tilde\p^\rho,\p^\rho,g_{ij}^\rho]=W_{\tau,\tau'}[\tilde\p,\p,\rho g_{ij}]=W_{\tau/\sqrt\rho,\tau'/\sqrt\rho}[\tilde\p,\p,g_{ij}],
\label{weyl}
\ee
which firstly shows that the functional evaluated for the scaled fields is the same
as the Weyl transformed functional, and secondly that this transformation can be absorbed into a rescaling of $\tau$ and $\tau'$. This implies that $W_{\tau,\tau'}[\tilde\p,\p]$ is analytic in the complex $\tau$-plane cut along the negative real axis. In particular, since the part that is quadratic in $\tilde\p$
is $\tau^{-d}\int d^d\x\tilde\p\Gamma(-\tau^2\nabla^2)\tilde\p$, it follows that
$\Gamma(-(\tau^2/\rho)\nabla^2)$ is analytic in the cut $\rho$-plane.
This allows us to express $\Gamma(-\tau^2/\nabla^2)$ for arbitrary
$\tau$ in terms of the local expansion (\ref{local})

\bq
&&\Gamma\left(-\tau^2\nabla^2\right)=\lim_{\lambda\rightarrow\infty}
{1\over 2\pi i}\int_C{d\rho\over \rho-1}\,e^{\lambda(\rho-1)}\,
\Gamma\left(-(\tau^2/\rho)\nabla^2\right)
\nonumber\\
&&\quad\quad
=\lim_{\lambda\rightarrow\infty}
\sum_{n=0}^\infty b_n
{1\over 2\pi i}\int_C{d\rho\over \rho-1}\,{e^{\lambda(\rho-1)}\over \rho^n}
(-\tau^2\nabla^2)^n\label{repo}
\eq
since the large contour, $C$, on which we can use the local expansion, (\ref{local}),
can be collapsed to a contribution from the cut, which is suppressed 
for large positive $\lambda$ and the pole at $\rho=1$ which gives us the 
left hand side.
Expanding the denominators in powers of $1/\rho$ gives, for example

\be
\Gamma\left(-\tau^2\nabla^2\right)=\lim_{\lambda\rightarrow\infty}
\sum_{n,r=0}^\infty (-)^r{b_n\,\lambda^{n+r}\,(-\tau^2\nabla^2)^n
\over(n-1)!
\,r!\,(n+r)}
\ee

The free energy is the  $\tau\rightarrow \infty,\,\tau'\rightarrow 0$ limit of $F[\tau',g_{ij}]=\int d^d\x \, a_{\tau,\tau'}$. A Weyl scaling of $g_{ij}$ can be compensated 
by scaling $\tau$, and $\tau'$ so when $\delta g_{ij}=g_{ij}\,\delta\rho$ the change in $F$ is

\be
\delta F=-{\delta\rho\over 2}\left(\tau{\partial F\over\partial\tau}+\tau'{\partial F\over\partial\tau'}\right)
\ee

$F$ satisfies
equations similar to regulated versions of (\ref{eqc}) (even if we include interactions), that follow from the \Sc equations (\ref{Schr}) and (\ref{Schr'}). If we use the
same regulator as before, cutting off the large eigenvalues of $\nabla^2$,
then
\be 
{\partial F\over\partial\tau}={1\over 2\tau}\int d^d\x \,
\left(\tau^{p+1}\Gamma_{\tau,\tau'}+{{p+1}\over 2}\right)\,\theta\left(s\tau^2\nabla^2+1\right)\,\delta^d(\x-\y)|_{\x=\y}.
\ee
and
\be 
-{\partial F\over\partial\tau'}={1\over 2\tau'}\int d^d\x \,
\left(\tau'^{p+1}\Upsilon_{\tau,\tau'}-{{p+1}\over 2}\right)\,\theta\left(s\tau'^2\nabla^2+1\right)\,\delta^d(\x-\y)|_{\x=\y}.
\ee
If we represent the step function by
\be
\label{step}
\theta( x)={1\over 2\pi i }\int_{C'} {dy\over y}\,e^{iyx},
\ee
with $C'$ a contour running just below the real axis, and if $f$ is some
function of $-\nabla^2$
then
\be
f(-\nabla^2)\,\theta\left(s\tau^2\nabla^2+1\right)\,\delta^d(\x-\y)
={1\over 2\pi i }\int_{C'} {dy\over y}e^{iy}\,f \left({i\over s\tau^2}{\partial\over\partial y}\right)
\,e^{iys\tau^2\nabla^2 }\,\delta^d(\x-\y).\label{ff}
\ee
Now  $e^{it\nabla^2 }\,\delta^d(\x-\y)\equiv H(t,\x,\y)$ satisfies the finite-dimensional
\Sc equation
\be
i{\partial\over\partial z}H=-\nabla^2 H,\quad H(0,\x,\y)=\delta^d(\x-\y).
\ee
At coincident
argument $H$ has the small $z$ expansion in powers of derivatives of $g_{ij}$ 
\be
H(z,\x,\x)\sim {\sqrt g\over (4\pi iz)^{d/2} }\sum_{n=0}^\infty a_n(\x)\,z^{n}.
\label{hexx}\ee
The $a_n(\x)$ are scalars made out of the metric and its derivatives at $\x$,
and $a_0(\x)=1$. They depend on the radius of the sphere as
$a_n\sim r^{-2n}$. Thus we can express (\ref{ff}) as

\be
\sum_{n=0}^\infty {1\over (4\pi i)^{d/2}}\left(\int {d^d\x\over \tau^d}
\sqrt g\,a_n(\x)\,\tau^{2n}\right){1\over 2\pi i}\int_{C'}{dy\over y}e^{iy}\,f \left({i\over s\tau^2}{\partial\over\partial y}\right)
\,(sy)^{n-d/2}
\label{delF}
\ee
Only a finite number of the $a_m$ survive as $r\rightarrow\infty$. For the case of a two-dimensional boundary these are just $a_0=1$ and 
$a_1=iR/6$, and the conformal anomaly is proportional to $a_1$. When the boundary has $2N$ dimensions the conformal anomaly is proportional to $a_N$.
If we assume that $f$ has a series expansion, $f(x)=\sum \tilde f_n x^n$ then we can compute the 
relevant integral in (\ref{delF}) as

\bq
&&
{1\over 2\pi i}\int_{C'}{dy\over y}e^{iy}\,f \left({i\over s\tau^2}{\partial\over\partial y}\right)
\,(sy)^{N-d/2}
\nonumber\\
&&
=\sum \tilde f_n\, s^{N-d/2}\left(-{i\over s\tau^2}\right)^n
{\sin(\pi d/2)\,(-1)^{N+1}\,(N-d/2)!\over \pi \,(N-d/2-n)}
\nonumber\\
&&
=\sin(\pi d/2)\,(-1)^{N}\,(N-d/2)! {1\over\pi } 
\int_s^\infty ds'\,s'^{N-1-d/2}\, f\left(-{i\over s'\tau^2}\right)
\eq
where we have taken $N<d/2$. If $f$ has a finite limit,$f_{\rm lim}$, as $d\downarrow 2N$
then, for $d$ close to $2N$ this becomes
\bq
&&
(d/2-N)\,\int_s^\infty ds'\,s'^{N-1-d/2}\, f\left(-{i\over s'\tau^2}\right)
\nonumber\\
&&
=s^{N-d/2}\,f\left(-{i\over s\tau^2}\right)+\int_s^\infty ds'\,s'^{N-d/2}\,{d\over ds'} f\left(-{i\over s'\tau^2}\right)
\eq
which tends to
$f_{\rm lim}(0)$ as $d\downarrow 2N$. Putting all this together we obtain the conformal anomaly as the large $\tau$ small $\tau'$ limit of

\be
\delta F=-{\delta\rho\over 4}\left(p+1+\hat\Gamma
-\hat\Upsilon
\right){1\over (4\pi i)^N}\int {d^d\x}
\sqrt g\,a_N(\x)
\ee
where
\bq
&&
\hat\Gamma=\lim_{\xi\rightarrow 0}\left(\lim_{d\downarrow 2N}\left(
\Gamma(-\tau^2\xi)+\left\{(\tau/\tau')^{p+1}\Upsilon(-\tau'^2\xi) -\Upsilon(-\tau^2\xi)\right\}^{-1} \,\Xi(-\tau^2\xi)^2\right)\right),
\nonumber\\&&
\hat\Upsilon=\lim_{\xi\rightarrow 0}\left(\lim_{d\downarrow 2N}\left(\left\{\Upsilon(-\tau'^2\xi)-({\tau'/\tau})^{p+1}\Upsilon(-\tau^2\xi)\right\}^{-1} \,\Xi(-\tau'^2\xi)^2-\Gamma(-\tau'^2\xi)\right)\right).
\eq
From the series expansions (\ref{localsol}) we see that
$\hat\Gamma=b_0=-(p+1)$ and for generic values of $p$ we have
$\hat\Upsilon=0$.
The order of the limits is important since when
$p$ approaches an odd integer as $d\downarrow 2N$
the coefficient $f_N$ diverges so that for $\xi\neq 0$ there is a suppression of 
$\left\{\Upsilon(-\tau'^2\xi)-({\tau'/\tau})^{p+1}\Upsilon(-\tau^2\xi)\right\}^{-1}$ thus $\hat\Upsilon=-\hat\Gamma$ which is just $-b_0$.
Since $p+1=\sqrt{d^2+4m^2}=2\sqrt{N^2+m^2}$ we have that when $\sqrt{N^2+m^2}$ is an integer, $\tilde N$,
the conformal anomaly is
\be
\delta F={\delta\rho\over 2}{\tilde N\over (4\pi i)^N}\int {d^{2N}\x}
\sqrt g\,a_N(\x)
\ee
otherwise it vanishes. In particular for ${\bf d=2}$ we have that
the central charge of the Virasoro algebra, $c$, equals $\tilde N$ when
$m=\sqrt{{\tilde N}^2-1}$ and vanishes otherwise.
For ${\bf d=4}$

\be
\delta F=-{\delta\rho\over 32\pi^2}\int d^4x\,\sqrt g\, a_2({\bf x})\,\tilde N
\ee
or zero, where
\be
a_2={1\over 180}\left(R_{ijkl}\,R^{ijkl}-R_{ij}\,R^{ij}-6\Box R+{5\over 2}R^2\right).
\label{kera}\ee
Our conventions for the curvature tensors are as in \cite{Birr}, 
i.e. $R^i_{\,\, jkl}=\partial_l\Gamma^i_{jk}-..$, $R_{ij}=R^k_{\,\,ikj}$ and
$\Box=g^{ij}\nabla_i\nabla_j$. The term in $\Box R$ can be removed by a counter-term
proportional to $R^2$.
The mass condition $m^2={\tilde N^2-N^2}$,
corresponds to the mass spectrum of the scalar fields
of Supergravity compactified on $AdS_5\times S^5$, \cite{Kim}, for $d=2N=4$,
and on $AdS_7\times S^4$ for $d=2N=6$, \cite{20}.
Note that for $D=2N\ge 4$ there are negative values of $m^2$ with non-vanishing
conformal anomaly.

\section{Example: free fermion field}

Another example is given by a free fermion theory. In order to describe this,
it will be necessary to discuss some subtleties arising from the representation
of fermion fields.

Wave functionals will be taken to be overlaps \( \langle u,u^{\dagger }|\Psi \rangle  \)
with a field state \( \langle u,u^{\dagger }| \), upon which fermion operators
act as follows. We diagonalize half of the components of the fermion fields
\begin{eqnarray}
\langle u,u^{\dagger }|Q_{+}\hat{\psi } & = & \sqrt{2}Q_{+}u\langle u,u^{\dagger }|\nonumber \\
\langle u,u^{\dagger }|\hat{\psi }^{\dagger }Q_{-} & = & \sqrt{2}u^{\dagger }Q_{-}\langle u,u^{\dagger }|,\label{rep1} 
\end{eqnarray}
where \( Q_{\pm }=\frac{1}{2}(1\pm Q) \) are for the moment arbitrary projection
operators. The other half are represented by functional differentiation
\begin{eqnarray}
\langle u,u^{\dagger }|Q_{-}\hat{\psi } & = & \frac{1}{\sqrt{2}}Q_{-}\frac{\delta }{\delta u^\dagger}\langle u,u^{\dagger }|\nonumber \\
\langle u,u^{\dagger }|\hat{\psi }^{\dagger }Q_{+} & = & \frac{1}{\sqrt{2}}\frac{\delta }{\delta u}Q_{+}\langle u,u^{\dagger }|.\label{rep2} 
\end{eqnarray}
 We can make the dependence of the field state on the Grassmann-valued source
fields explicit by writing

\begin{equation}
\label{dep}
\langle u,u^{\dagger }|=\langle Q|\exp {\sqrt 2}\,{\mathrm{tr}}\int d^{d}{x}(u^{\dagger }Q_{-}\hat{\psi }-\hat{\psi }^{\dagger }Q_{+}u),
\end{equation}
where \( \langle Q| \) is defined by
\begin{equation}
\label{defq}
\langle Q|Q_{+}\hat{\psi }=\langle Q|\hat{\psi }^{\dagger }Q_{-}=0.
\end{equation}
Thus the choice of \( Q \) corresponds to a choice of Dirac sea, though \emph{not}
a physical Dirac sea; these constraints are merely artifacts of our choice of
representation. On the other hand this choice should not be considered completely
arbitrary; since the field-states parametrized by \( \langle Q| \) are non-physical,
we must be careful that their overlaps with physical states are well-defined
and non-vanishing. Also, we may wish to include gauge interactions; if we want
wave-functionals to be invariant under local gauge transformations, we find
that we must choose \( Q \) to be a local, field independent operator. In particular,
if we choose \( Q_{\pm } \) to be projectors onto +ve/-ve energy eigenstates,
the resulting wave-functionals do \emph{not} satisfy Gauss' law \cite{1},\cite{2}. For reasons that will become clear, in \( AdS \) space we will choose \( Q=\pm \gamma ^{0} \).

Now define

\begin{equation}
\label{norm}
\langle\!\langle u,u^{\dagger }|\equiv \langle u,u^{\dagger }|e^{{\mathrm{tr}}\int d^{d}x[u^{\dagger }Qu]}=\langle Q|e^{\sqrt{2}{\mathrm{tr}}\int d^{d}x(u^{\dagger }\hat{\psi }-\hat{\psi }^{\dagger }u)},
\end{equation}
using which it may be verified that our representation coincides with that given
in \cite{1}. 

\begin{eqnarray}
\langle\!\langle u,u^{\dagger }|\hat{\psi } & = & \frac{1}{\sqrt{2}}(u+\frac{\delta }{\delta u^{\dagger }})\langle\!\langle u,u^{\dagger }|\nonumber \\
\langle\!\langle u,u^{\dagger }|\hat{\psi }^{\dagger } & = & \frac{1}{\sqrt{2}}(u^{\dagger }+\frac{\delta }{\delta u})\langle\!\langle u,u^{\dagger }|.\label{jackiw} 
\end{eqnarray}
Similarly, if we define 

\begin{equation}
\label{othernorm}
|{v},{v}^{\dagger }\rangle\!\rangle =e^{\sqrt{2}{\mathrm{tr}}\int d^{d}x({v}^{\dagger }\hat{\psi }-\hat{\psi }^{\dagger }{v})}|Q\rangle ,
\end{equation}
with \( Q_{-}\hat{\psi }|Q\rangle =\hat{\psi }^{\dagger }Q_{+}|Q\rangle =0 \),
then we have

\begin{eqnarray}
\hat{\psi }|{v},{v}^{\dagger }\rangle\!\rangle  & = & \frac{1}{\sqrt{2}}({v}-\frac{\delta }{\delta {v}^{\dagger }})|{v},{v}^{\dagger }\rangle\!\rangle \nonumber \\
\hat{\psi }^{\dagger }|{v},{v}^{\dagger }\rangle\!\rangle  & = & \frac{1}{\sqrt{2}}({v}^{\dagger }-\frac{\delta }{\delta {v}})|{v},{v}^{\dagger }\rangle\!\rangle .\label{otherrep} 
\end{eqnarray}

Written in the form (\ref{jackiw}) or (\ref{otherrep}), the representation
is reducible, but we are free to take 
\begin{equation}
\label{constraints}
Q_{-}u=u^{\dagger }Q_{+}=0,
\end{equation}
and
\begin{equation}
\label{otherconstraints}
Q_{+}{v}={v}^{\dagger }Q_{-}=0,
\end{equation}
which removes the reducibility. To avoid taking functional derivatives with
respect to constrained fields, however, we find it most convenient to work with
the unconstrained sources, and impose the constraints at the end of our calculations.
The states \( |{v},{v}^{\dagger }\rangle  \) and \( \langle u,u^{\dagger }| \)
depend only on the unconstrained components in any case, and the relative factor
of \( e^{{\mathrm{tr}}\int d^{d}x[u^{\dagger }Qu]} \) appearing in \( \langle\!\langle u,u^{\dagger }| \)
merely ensures that we stay in the appropriate Fock space when we apply field
operators in accordance with (\ref{jackiw}) and (\ref{otherrep}). 

Imposing the constraints will nevertheless be of physical importance; for example
in the \( AdS \)/CFT correspondence they cause a Dirac fermion in the \( AdS \)
theory to become a chiral fermion in the boundary theory. This is in agreement
with the findings of other authors, for example in \cite{3}.

Now we are interested in the Schr\"odinger functional
\begin{equation}
\label{fun1}
\Psi _{\tau ,\tau ^{\prime }}[u,u^{\dagger },{v},{v}^{\dagger }]=\langle\!\langle u,u^{\dagger }|T\exp (-\int ^{\tau }_{\tau \prime }dt\hat{H}(t))|{v},{v}^{\dagger }\rangle\!\rangle ,
\end{equation}
so we need to find the overlap \( \langle\!\langle u,u^{\dagger }|{v},{v}^{\dagger }\rangle\!\rangle  \)
which is the coincident time limit of the above. The definitions (\ref{defq})
and (\ref{norm}) give rise to the following equations:
\begin{eqnarray}
0 & = & (\hat{\psi }^{\dagger }-\sqrt{2}u^{\dagger })Q_-\langle\!\langle u,u^{\dagger }|{v},{v}^{\dagger }\rangle\!\rangle \nonumber \\
 & = & Q_{-}(\hat{\psi }+\sqrt{2}{v})\langle\!\langle u,u^{\dagger }|{v},{v}^{\dagger }\rangle\!\rangle \nonumber \\
 & = & (\hat{\psi }^{\dagger }+\sqrt{2}{v}^{\dagger })Q_+\langle\!\langle u,u^{\dagger }|{v},{v}^{\dagger }\rangle\!\rangle \nonumber \\
 & = & Q_{+}(\hat{\psi }-\sqrt{2}u)\langle\!\langle u,u^{\dagger }|{v},{v}^{\dagger }\rangle\!\rangle ,\label{e4} 
\end{eqnarray}
where we have used the canonical commutation relations. The field operators
in (\ref{e4}) may be represented by either pair of source fields, and the equations
solved to give

\begin{equation}
\label{bound}
\langle\!\langle u,u^{\dagger }|{v},{v}^{\dagger }\rangle\!\rangle =\exp {\mathrm{tr}}\int d^{d}{x}(u^{\dagger }Qu-u^{\dagger }2Q_{-}{v}+{v}^{\dagger }2Q_{+}u+{v}^{\dagger }Q{v}).
\end{equation}

We are now ready to consider the AdS metric (\ref{met}). With the choice of vielbein
\begin{equation}
\label{vie}
e^{a}_{\mu }=t^{-1}\delta ^{a}_{\mu },
\end{equation}
the Euclidean action is given by\footnote{
Gamma matrices obey \( \{\gamma ^{i},\gamma ^{j}\}=2\delta ^{ij} \) throughout
this section.
} 

\begin{equation}
\label{act}
S=\int d^{d+1}x\sqrt{g}\bar{\psi }(t\gamma \cdot D-m)\psi =\int d^{d+1}xt^{-d}\bar{\psi }(\gamma \cdot \partial -\frac{\gamma ^{0}d}{2t}-\frac{m}{t})\psi ,
\end{equation}
since according to (\ref{met}), \( \sqrt{g}=t^{-d-1} \), and the spin covariant derivative
is \( D_{\mu }=\partial _{\mu }-\frac{1}{t}\Sigma _{0\mu } \). Changing
variables to
\( \phi =t^{-d/2}\psi  \) and \( \phi ^{\dagger }=t^{-d/2}\psi ^{\dagger } \)
the action becomes
\begin{equation}
S = \int d^{d+1}x\bar{\phi }(\gamma \cdot \partial -\frac{m}{t})\phi ,
\end{equation}
and  for \( m=0 \) it coincides with the flat-space
action. As in the bosonic case, if we also put $\bar t=\ln t$, the volume
element in the corresponding path-integral becomes the usual flat-space one
induced by $||\delta\phi||^2=
\int d\bar t\,d\x \,\delta\phi^\dagger\delta\phi$. Thus we can make use of the representation
(\ref{jackiw}) for $\phi$ and $\phi^\dagger$.
The integrands
in (\ref{dep}) and (\ref{bound}) do not aquire a factor from the metric, as
this has been absorbed into the definition of the fields. 

The partition function is again given by the \( \tau ^{\prime }\rightarrow 0 \),
\( \tau \rightarrow \infty  \) limit of the Schr\"odinger functional, with
\( {u}={u}^{\dagger }=\lim _{|{{\mathbf{x}}}|\rightarrow \infty }v({{\mathbf{x}}})=\lim _{|{{\mathbf{x}}}|\rightarrow \infty }v^{\dagger }({\mathbf{x}})=0 \). 
In path-integral form the \Sc functional is

\begin{equation}
\label{part}
\Psi _{\tau ,\tau ^{\prime }}[u,u^{\dagger },{v},{v}^{\dagger }]=\int {\cal D}\phi {\cal D}\phi ^{\dagger }e^{-S-S_B}
\end{equation}
where the boundary term is
\bq
S_B=&&\int_{x^0=\tau'} d^dx\left(\phi^\dagger Q_-\phi-{\sqrt 2}\phi^\dagger Q_-v
+{\sqrt 2} v^\dagger Q_+\phi\right)\nonumber \\&&
-\int_{x^0=\tau} d^dx\left(\phi^\dagger Q_+\phi-{\sqrt 2}\phi^\dagger Q_+u
+{\sqrt 2} u^\dagger Q_-\phi\right).
\eq
If $\phi$ and $\phi^\dagger$ are integrated over freely then we can shift
them by solutions to the classical equations of motion. Choosing these solutions
to satisfy the 
boundary conditions corresponding to (\ref{rep1})
\begin{eqnarray}
t=\tau':&&\quad Q_{-}\phi   = -\sqrt{2}Q_{-}v,
\quad\phi ^{\dagger }Q_{+}  =  -\sqrt{2}v^{\dagger }Q_{+}\nonumber \label{pathconds1} \\
t=\tau:&&\quad Q_+\phi   =  {\sqrt 2}u,\quad\phi^\dagger Q_-={\sqrt 2}u^\dagger
Q_-
\end{eqnarray}
causes the action to separate into a piece depending only on the integration
variables and a piece depending only on the classical solution.
Our boundary term $S_B$ is thus determined by the conditions
(\ref{rep1}).
Note that the
classical action does \emph{not} vanish, and there is therefore no need to add
any additional boundary term with undetermined coefficients, as in \cite{3}.
(Other authors have discussed boundary terms for fermions \cite{11}-\cite{12}).

The Schr\"odinger equation is
\begin{equation}
\label{sch}
-\frac{\partial }{\partial \tau }\Psi _{\tau ,\tau ^{\prime }}=\hat{H}\Psi _{\tau ,\tau ^{\prime }},
\end{equation}
where
\begin{equation}
\label{ham}
\hat{H}=\frac{1}{2}\int d^{d}x(u^{\dagger }+\frac{\delta }{\delta u})h(u+\frac{\delta }{\delta u^{\dagger }}),\qquad h=(\gamma ^{0}\gamma ^{i}\partial _{i}-\frac{\gamma ^{0}m}{\tau }).
\end{equation}

Now the logarithm of the partition function is obtained as \( \lim _{\tau \rightarrow \infty }W_{\tau ,0}[0,0,{v},{v}^{\dagger }] \),
where \( W_{\tau ,0}[u,u^{\dagger },{v},{v}^{\dagger }]=\log \Psi _{\tau ,0}[u,u^{\dagger },{v},{v}^{\dagger }] \)
may be expanded in analogy with (\ref{expand}) as

\begin{equation}
\label{log}
\int d^{d}x\left\{ f+u^{\dagger }\Gamma (\tau \gamma _{0}\gamma ^{i}\partial _{i})u+u^{\dagger }\Xi (\tau \gamma _{0}\gamma ^{i}\partial _{i}){v}+{v}^{\dagger }\Pi (\tau \gamma _{0}\gamma ^{i}\partial _{i})u+{v}^{\dagger }  \Upsilon (\tau \gamma _{0}\gamma ^{i}\partial _{i}){v}\right\} .
\end{equation}

Substituting (\ref{log}) into (\ref{sch}) gives
\begin{equation}
\label{eqs}
\left\{ \begin{array}{lr}
\dot{\Gamma }=-\frac{1}{2}(1-\Gamma )h(1+\Gamma ) & \\
\dot{\Xi }=-\frac{1}{2}(1-\Gamma )h\Xi  & \qquad \Gamma (0)=\Upsilon (0)=Q\\
\dot{\Pi }=\frac{1}{2}\Pi h(1+\Gamma ) & \Xi (0)\sim-2Q_{-}\\
\dot{\Upsilon }=\frac{1}{2}\Pi h\Xi  & \Pi (0)\sim2Q_{+}\\
\dot{f}=\frac{1}{2}{\mathrm{Tr}}h(1+\Gamma )\delta ^{d}(\x)|_{{\mathbf{x}}=0} & 
\end{array}\right. 
\end{equation}
where the initial conditions are read off from (\ref{bound}). We will find that
$\Xi(0)$ and $\Pi(0)$ diverge as a result of the ill-defined nature of
$\lim_{\tau\rightarrow 0}\int^\tau_0{m\over t}dt$, but that
$\lim_{\tau\rightarrow\tau^\prime}\Psi_{\tau,\tau^\prime}$ is well-defined
and given by (\ref{bound}) for all $\tau^\prime\ne 0$.

As may be verified by direct substitution, the equation for \( \Gamma  \) is
solved by
\begin{equation}
\label{ss}
\Gamma =(\Sigma -Q_{-})(\Sigma +Q_{-})^{-1},
\end{equation}
with \( \Sigma  \) satisfying the Dirac equation \( \dot{\Sigma }+h\Sigma =0 \).
Making the ansatz
\begin{equation}
\label{anz}
\Sigma =\sum _{n=0}^{\infty }a_{n}(-\tau \gamma ^{0}\gamma ^{i}\partial _{i})^{n}Q_{+}\tau ^{-m},
\end{equation}
leads (for the specific choice \( Q=\gamma ^{0} \)) to the recurrence relation
\( a_{n}=\frac{a_{n-1}}{n+m(1-(-1)^{n})} \). The boundary condition is satisfied
if \( a_{0}=1 \) and \( m\ge 0 \), and we can explicitly sum the series in terms
of Bessel functions. In momentum space
\begin{eqnarray}
\Sigma  & = & \Gamma (1/2+m)\left( \frac{{p}\tau }{2}\right) ^{1/2-m}(I_{m-1/2}({p}\tau )+PI_{m+1/2}({p}\tau ))Q_{+}\tau ^{m}\nonumber \\
 & = & (E+O)Q_{+}.\label{sigma} 
\end{eqnarray}
Here \( P= \)\( \frac{i\gamma ^{0}\gamma \cdot \mathbf{p}}{|\mathbf{p}|} \);
the operators \( \frac{1}{2}(1\pm P) \) project onto +ve/-ve eigenvalues of
the massless flat-space hamiltonian \( \gamma ^{0}\gamma ^{i}\partial _{i} \). 
Substituting back into (\ref{ss}) we find that
\begin{eqnarray}
\Gamma  & = & Q+2Q_{-}OE^{-1}\nonumber \\
 & = & Q+2Q_{-}P\frac{I_{m+1/2}({p}\tau )}{I_{m-1/2}({p}\tau )}.\label{gamma} 
\end{eqnarray}
Next, as may again be verified by substitution, the equation for \( \Pi  \)
has the solution
\begin{eqnarray}
\Pi  & = & 2Q_{+}(\Sigma +Q_{-})^{-1}\nonumber \\
 & = & 2Q_{+}E^{-1},\label{pi} 
\end{eqnarray}
which gives the expected divergent behaviour at \( \tau =0 \).

Now consider the equation for \( \Xi  \). We can rewrite \( \Gamma  \) in
the following way: \\
\( \Gamma =-(\bar{\Sigma }+Q_{+})^{-1}(\bar{\Sigma }-Q_{+}) \) where 
\begin{equation}
\label{altsols}
\bar{\Sigma }=Q_{+}(E-O).
\end{equation}
which satisfies \( \dot{\bar{\Sigma }}-\bar{\Sigma }h=0 \). This enables us to find
the solution
\begin{eqnarray}
\Xi  & = & -2(\bar{\Sigma }+Q_{-})^{-1}Q_{-}\nonumber \\
 & = & -2Q_{-}E^{-1}.\label{xi} 
\end{eqnarray}

Finally, to solve the equation for \( \Upsilon  \) we put \( \Upsilon =\Pi R\Xi +Q \)
where \( R \) satisfies \( \{R,Q\}=0 \). Substituting into (\ref{eqs}) gives
\begin{equation}
\label{reqn}
\Pi \left( 2\dot{R}-(1+4ROE^{-1})\gamma ^{0}\gamma ^{i}\partial _{i}\right) \Xi =0.
\end{equation}
 Now define \( \tilde{O} \), \( \tilde{E} \) such that \( O\rightarrow \tilde{O} \),
\( E\rightarrow \tilde{E} \) as \( m\rightarrow -m \). The correponding expansion coefficients $\tilde a_n$ are divergent as $m-1/2$ approaches an integer, so we keep $m$ variable, allowing it to approach such values within convergent
expressions. This is the analogue of keeping $d$ variable in the scalar case.
The necessity for such regularisation is due to our working in momentum
space, rather than configuration space. 
Using the identity
\( E\tilde{E}-O\tilde{O}=1 \), we find that \( R=-\frac{1}{2}\tilde{O}E \),
so that
\begin{eqnarray}
\Upsilon  & = & Q+2Q_{+}\tilde{O}E^{-1}\nonumber \\
 & = & Q+2CQ_{+}P{p}^{2m}\frac{I_{1/2-m}({p}\tau )}{I_{m-1/2}({p}\tau )},\label{upsilon} 
\end{eqnarray}
where \( C= \)\( 2^{-2m}\frac{(-1/2-m)!}{ (m-1/2)!} \). This is
nonlocal even when $m=n+1/2$ for some integer $n$ causing the Bessel functions to cancel because
then $C$ diverges and we have a
similar situation to that which we encountered in the bosonic case.

The large \( \tau  \) behaviour of \( \Upsilon  \) is easily found: 
\begin{equation}
\label{limt}
\lim _{\tau \rightarrow \infty }\Upsilon =2CQ_{+}P{p}^{2m}.
\end{equation}
Fourier transforming, we find that the partition function is given by
\begin{equation}
\label{partition}
\log Z[{v},{v}^{\dagger }]=\int d^{d}{x}d^{d}{y}{v}^{\dagger }({\mathbf{x}})\left( K\gamma ^{0}\frac{\gamma \cdot ({\mathbf{x}}-{\mathbf{y}})}{|{\mathbf{x}}-{\mathbf{y}}|^{d+2m+1}}\right) {v}({\mathbf{y}}),
\end{equation}
where \( K=-\frac{ 2(\frac{d-1}{2}+m)!}{ (m-1/2)!\pi ^{d/2}} \).
Here we have imposed the constraints (\ref{constraints}). This is the correct
two-point function for a quasi-primary fermion field of scaling dimension \( d/2+m \). 

Now all this assumed that \( m\ge 0 \); but if $m<0$ the above argument holds with \( m \)
replaced by \( -m \) provided we take \( Q=-\gamma ^{0} \). Thus we conclude
that the scaling dimension is \( d/2+|m| \) in general.

An incidental point of interest is the relationship of our solution of the Schr\"odinger
equation to the classical field configuration \( \phi  \) to which it corresponds.
This is not quite the same as the solution \( \Sigma  \) to the Dirac equation
which we found, because the latter is divergent as \( \tau \rightarrow \infty  \).
However, defining \( \tilde{\Sigma } \) by \( \Sigma \rightarrow \tilde{\Sigma } \)
as \( m\rightarrow -m \), we find that \( P\tilde{\Sigma } \) also solves
the Dirac equation. Taking a suitable linear combination of these two solutions
(and allowing them to operate on the boundary value \( {v} \)) we can
construct a unique field configuration which satisfies the appropriate boundary
conditions and is finite as \( \tau \rightarrow \infty  \). By Fourier transforming
this configuration, we found that it coincides exactly with that given in
\cite{3} (when we change back to the variables in the original action).

Now that we have found $\Psi_{\tau,0}$, we can construct
$\Psi_{\tau,\tau^\prime}$ from the self-reproducing property. The
inner-product of wave-functionals follows from (\ref{bound}): 
\begin{eqnarray}
\langle 1|2\rangle &=& \int DuDu^\dagger DvDv^\dagger \langle
1|u,u^\dagger\rangle\langle u,u^\dagger|v,v^\dagger\rangle\langle
v,v^\dagger|2\rangle\nonumber\\ &=&  \int DuDu^\dagger DvDv^\dagger \langle
1|u,u^\dagger\rangle\langle v,v^\dagger|2\rangle e^{2v^\dagger u-2u^\dagger
  v}.
\end{eqnarray}
It is
important to note that we integrate over the {\em constrained\/} fields,
(\ref{constraints}),
which reflect the true functional dependence of the wave-functionals. This
allows us to drop the $Q$-dependence from $\Gamma$, etc. so we write the
logarithm of $\Psi_{\tau,\tau^\prime}$ as
\be
\int d^dx\left\{f_{\tau,\tau^\prime}+u^{\dagger
    }\Gamma_{\tau,\tau^\prime}u+u^{\dagger
    }\Xi_{\tau,\tau^\prime}{v}-{v}^{\dagger
    }\Xi_{\tau,\tau^\prime}u+{v}^{\dagger }
  \Upsilon_{\tau,\tau^\prime}{v}\right\},
\ee
and we have $\Gamma_{\tau,0}=2OE^{-1}$,
$\Xi_{\tau,0}=-2E^{-1}$, and
$\Upsilon_{\tau,0}=2\tilde OE^{-1}$. Note that
$\Pi_{\tau,0}=-\Xi_{\tau,0}$. Then from
\be
\Psi_{\tau,0}[\tilde u,\tilde u^\dagger,\tilde v,\tilde v^\dagger]=\int d^dx
DuDu^\dagger DvDv^\dagger\Psi_{\tau,\tau^\prime}[\tilde u,\tilde
u^\dagger,u,u^\dagger]\Psi_{\tau^\prime,0}[v,v^\dagger,\tilde v,\tilde
v^\dagger]e^{2v^\dagger u-2u^\dagger
  v},
\ee
we obtain
\begin{eqnarray}
\Gamma_{\tau,0} &=&
\Gamma_{\tau,\tau^\prime}+\Xi_{\tau,\tau^\prime}^2\Upsilon_{\tau,\tau^\prime}^{-1}-4\Xi_{\tau,\tau^\prime}^2\Upsilon_{\tau,\tau^\prime}^{-2}(4\Upsilon_{\tau,\tau^\prime}^{-1}+\Gamma_{\tau^\prime,0})^{-1}\nonumber\\
\Xi_{\tau,0} &=&
-2\Xi_{\tau,\tau^\prime}\Xi_{\tau^\prime,0}\Upsilon_{\tau,\tau^\prime}^{-1}(4\Upsilon_{\tau,\tau^\prime}^{-1}+\Gamma_{\tau^\prime,0})^{-1}\nonumber\\
\Upsilon_{\tau,0} &=&
\Upsilon_{\tau^\prime,0}+\Xi_{\tau^\prime,0}^2(4\Upsilon_{\tau,\tau^\prime}^{-1}+\Gamma_{\tau,0})^{-1}
\end{eqnarray}
and hence
\begin{eqnarray}
\Upsilon_{\tau,\tau^\prime} &=& -4(\Xi_{\tau^\prime,0}^2(\Upsilon_{\tau^\prime,0}-\Upsilon_{\tau,0})^{-1}+\Gamma_{\tau^\prime,0})^{-1}\nonumber\\
\Xi_{\tau,\tau^\prime} &=& {1\over2}\Xi_{\tau^\prime,0}\Pi_{\tau,0}(\Upsilon_{\tau^\prime,0}-\Upsilon_{\tau,0})^{-1}\Upsilon_{\tau,\tau^\prime}\nonumber\\
\Gamma_{\tau,\tau^\prime} &=& \Gamma_{\tau,0}-{1\over4}\Xi_{\tau^\prime,0}^2\Xi_{\tau,0}^2(\Upsilon_{\tau^\prime,0}-\Upsilon_{\tau,0})^{-2}\Upsilon_{\tau,\tau^\prime}-(\Upsilon_{\tau^\prime,0}-\Upsilon_{\tau,0})^{-1}\Xi_{\tau,0}^2.
\end{eqnarray}
From this we can check that as $\tau\rightarrow\tau'\neq 0$ the \Sc functional
$\Psi_{\tau,\tau^\prime}$ reduces to $\langle u,u^\dagger|v,v^\dagger\rangle$
as it should.

We can calculate the conformal anomaly in the same way as before. Working
with the metric (\ref{newmet}) the Schr\"odinger equation is unchanged,
except that derivatives become covariant with respect to \( g \) and the Hamiltonian
density aquires a factor of \( \sqrt{g} \). We need to introduce a UV regulator,
which we would like to write in terms of a heat-kernel expansion, so it is convenient
to re-express everything in terms of positive definite operators. Hence, for example, we rewrite
the solution (\ref{gamma}) as
\begin{equation}
\label{re}
\Gamma (\tau \gamma ^{0}\gamma ^{i}\nabla _{i})=Q+2Q_{-}\tau \gamma ^{0}\gamma ^{i}\nabla _{i}\sum ^{\infty }_{n=0}d_{n}(\tau ^{2}D^{2})^{n},
\end{equation}
for some appropriate coefficients \( d_{n} \). We have defined \( D\equiv \gamma ^{0}\gamma ^{i}\nabla _{i}) \).

As before, a Weyl transformation may be implemented by scaling \( \tau  \),
so when \( \delta g_{ij}=g_{ij}\delta \rho  \) the free energy changes by
\begin{equation}
\label{anom}
\delta F=-\frac{\delta \rho }{2} \int d^{d}{x}\left(\tau\frac{\partial f_{\tau,\tau^\prime}}{\partial \tau }+\tau^\prime\frac{\partial f_{\tau,\tau^\prime}}{\partial \tau^\prime }\right).
\end{equation}

The Schr\"odinger equations yield equations for $F$ corresponding to regulated versions of (\ref{eqs})
\begin{equation}
\label{free}
\frac{\partial f}{\partial \tau }=\frac{1}{2}{\mathrm{tr}}\left(h(1+Q+Q_-\Gamma_{\tau,\tau^\prime})\theta (1-s\tau ^{2}D^{2})\delta ^{d}({\mathbf{x}}-\y)|_{{\mathbf{x}}=\y}\right),
\end{equation}
and
\begin{equation}
\label{free2}
-\frac{\partial f}{\partial \tau^\prime
  }=\frac{1}{2}{\mathrm{tr}}\left(h(1+Q+Q_+\Upsilon_{\tau,\tau^\prime} )\theta (1-s\tau ^{\prime2}D^{2})\delta ^{d}({\mathbf{x}}-\y)|_{{\mathbf{x}}=\y}\right).
\end{equation}

Representing the step function by (\ref{step}), we have
\begin{equation}
\label{reg}
\theta (1-s\tau ^{2}D^{2})\delta ^{d}({\mathbf{x}}-\y)\,1=\frac{1}{2\pi i}\int _{C^{\prime }}\frac{dy}{y}e^{iy}e^{-iys\tau ^{2}D^{2}}\delta ^{d}({\mathbf{x}}-\y)\,1,
\end{equation}
where \( e^{-izD^{2}}\delta ^{d}({\mathbf{x}}-{\mathbf{y}})\,1\equiv H(z,{\mathbf{x}},{{\mathbf{y}}}) \)
satisfies

\begin{equation}
\label{schro}
i\frac{\partial }{\partial z}H=D^{2}H,\qquad H(0,{\mathbf{x}},\y)=\delta ^{d}({\mathbf{x}}-\y)\,1,
\end{equation}
and has the small \( z \) expansion
\begin{equation}
\label{heat}
H(z,{\mathbf{x}},{\mathbf{x}})\sim \frac{\sqrt{g}}{(4\pi iz)^{d/2}}\sum _{n=0}^{\infty }\tilde a_{n}({\mathbf{x}})z^{n}.
\end{equation}

Thus a general function $g(D^2)$ satisfies
\begin{eqnarray}
\label{expg}
& & {\rm tr}\left(g(D^2)\theta (1-s\tau ^{2}D^{2})\delta
  ^{d}({\mathbf{x}}-\y)\right)={1\over 2\pi i }\int_{C'} {dy\over
  y}e^{iy}\,{\rm tr}\left(g \left({i\over s\tau^2}{\partial\over\partial y}\right)
\,e^{iys\tau^2D^2 }\,\delta^d(\x-\y)\right)\nonumber\\
& &\quad = {\rm tr}\left(\sum_{n=0}^\infty {1\over (4\pi i)^{d/2}}\left(\int {d^d\x\over \tau^d}
\sqrt g\,\tilde a_n(\x)\,\tau^{2n}\right){1\over 2\pi i}\int_{C'}{dy\over y}e^{iy}\,g \left({i\over s\tau^2}{\partial\over\partial y}\right)\right)
\,(sy)^{n-d/2}
\end{eqnarray}
where only terms up to $n=N$ contribute when $d=2N$. The conformal anomaly is
again proportional to $\tilde a_N$; in the limit $\tau\rightarrow\infty$,
$\tau^\prime\rightarrow0$ the other terms arising from (\ref{free}) and
(\ref{free2}) either vanish or reproduce ultraviolet divergences associated
with the renormalization of the cosmological constant etc. Using the same argument as in the scalar case, the term proportional to $\tilde a_N$ is

\be
\lim_{\xi\rightarrow 0}\left(g(\xi){1\over (4\pi i)^{N}}\int {d^d\x}
\sqrt g\,\tilde a_n(\x)\right)
\ee
For generic values of the mass $m$ the conformal anomaly can be computed from 
expansions of $\Gamma_{\tau,\tau^\prime}$ and
$\Upsilon_{\tau,\tau^\prime}$ in powers of $D$.
These expansions both begin with terms of order $D$.
Using these in (\ref{free}) and (\ref{free2}) gives
vanishing contributions to the anomaly.
There are also contributions from
${\rm tr}hQ$ which cancel between (\ref{free}) and (\ref{free2}).
So for generic values of the mass the conformal anomaly is zero.
But due to the divergent
nature of $\Upsilon_{\tau,0}$ as $2|m|\rightarrow 2\tilde N-1$ for any
positive integer $\tilde N$, we
have $\Upsilon_{\tau,\tau^\prime}\rightarrow -4\Gamma^{-1}_{\tau^\prime,0}$,
and this has a leading order term proportional to 
$1/D$ which combines with the $D$ in $h$ to give a finite
contribution as $\xi\rightarrow 0$. We conclude that the conformal anomaly is zero unless $2|m|$ is an odd integer
$2\tilde N-1$. For a four-dimensional boundary these are precisely the values appearing in the mass spectrum
of Supergravity compactified on $AdS_5\times S^5$ \cite{Kim},
and for a six-dimensional boundary they coincide with the mass spectrum
of Supergravity compactified on $AdS_7\times S^4$ \cite{20}. In this case we have
\be
\delta F=-\delta\rho\tilde N{1\over(4\pi i)^N}{\rm tr}\int d^{2N}x\sqrt
g Q_-a_N.
\ee

For \({\bf d=2 }\) and \( r\rightarrow \infty  \) the anomaly is proportional
to \( \tilde a_{1}=-iR\, 1/12 \), so for a fermion with \( \sigma  \) spinor components,

\begin{equation}
\label{ans}
\delta F=\frac{\delta \rho }{96\pi }\int d^{2}x\sqrt{g}R\tilde N\sigma,
\end{equation}
from which we identify the central charge as $\tilde N$.

For \( {\bf d=4} \) we have
\begin{equation}
\label{d4}
\delta F=\frac{\delta \rho }{64\pi^2 }\int d^{4}x\sqrt{g}{\rm tr}\tilde
a_{2}(\x)\tilde N ,
\end{equation}
where
\be
{\rm tr}\tilde a_2={\sigma\over 720}\left({5\over 2}R^2+6\Box R-{7\over 2}R_{ijkl}\,R^{ijkl}
-4R_{ij}\,R^{ij}\right).
\ee

\section{Gravitational Sector}

The tree-level conformal anomaly of pure gravity has been calculated for $AdS_3$
using the Chern-Simons formulation, \cite{Brown}-\cite{Banados}. In higher dimensions it has been computed by solving the Einstein equations perturbatively in terms of boundary data, \cite{Henningson}. We will now show how our \Sc functional technique
reproduces these results in a simple fashion. Our method employs a somewhat different regularization procedure, and so it is important to show that it is consistent with
earlier calculations. The main difference is that authors of \cite{Henningson} effectively compute the classical free energy of the gravitational sector by finding $W_{\tau,\tau'}$ for $\tau=\infty$ and $\tau'$ a small regulator,
whereas we take $\tau'=0$ and treat $\tau$ as a large regulator. (Since we
will work at tree-level it is not necessary to keep $\tau'$ non-zero
as we did in the earlier computations of one-loop anomalies).

\medskip
Formally we need to compute the Euclidean functional integral that represents a state of pure gravity, with Einstein-Hilbert action and cosmological constant $\Lambda<0$,
\be
Z[g_{rs}]=\int {\cal D}  G \, \exp \left(-\int d^{d+1} x\,
{\sqrt G} (R+2\Lambda)+{\rm boundary\, terms} \right)
\ee
where we should integrate over all metrics $G_{\mu\nu}$ of a $d+1$ dimensional manifold
which induce the metric $g_{rs}$ on the
boundary. This is ill-defined for a variety of reasons, such as non-renormalizability
and unboundedness of the action, which we ignore in the hope that these pathologies are
absent from the more fundamental theory of which this is only a part.
The integral over all metrics includes reparametrizations which can be factored out 
using the Faddeev-Popov method. The standard ADM decomposition of the metric is

\be
(G_{\mu\nu})=\left(
\begin{array}{cc}
N^2+N_iN_j G^{ij} & N_j\\
N_i & G_{ij}\end{array}
\right)
\ee
with $G^{ij}$ the inverse of the $d\times d$ matrix $G_{ij}$.
We will fix the gauge by choosing
\be
N^2=L^2/t^2,\quad N_i=0, \quad G_{ij}={g_{ij}+h_{ij}\over t^2}\label{ADM1}
\ee
with $t=x^0$. The dynamical variables are just the $h_{ij}$, and we take the boundary
to be at $t=\tau=0$, where $h_{ij}=0$. The gauge
conditions should be accompanied by the introduction of ghosts, but these will not contribute to the tree-level conformal anomaly. Expanding the action in powers of $h_{ij}$, and taking $L^2=-d(d-1)/(2\Lambda)$ gives

\bq&&
\int d^{d+1} x\,
{\sqrt G} (R+2\Lambda)-{\rm boundary\, terms}
=\int d^{d+1} x\,
{{\sqrt g}L\over t^{d-1}}\Big({d(d-1)\over L^2t^2}+R(g)
\nonumber\\
&&\quad\quad\quad
+h_{ij}\,\tilde G^{ij}(g)+(\dot h_{ij}\,\dot h^{ij}-(\dot h^i_i)^2)/(4L^2)+h_{ij}\,
\Box^{ijkl}h_{kl}+..\Big)\label{fiftwo}
\eq
where the boundary terms are chosen to make the action quadratic in first derivatives of $h_{ij}$, and the dots denote terms of higher order in $h_{ij}$.
Indices are raised and lowered with $g_{ij}$.
$R(g)$ is the d-dimensional curvature calculated by taking $g_{ij}$ as metric
and $\sqrt g(R(g)+h_{ij}\,\tilde G{ij}+h_{ij}\,
\Box^{ijkl}h_{kl}$ are the first three terms in the expansion of 
$\sqrt {det (g+h)}\,R(g+h)$, so that $\Box$ is a second order differential operator. The terms of higher order in $h$ each contain
one or two derivatives. As in section (1) the state $Z[g_{ij}]$ is the 
$\tau\rightarrow\infty$ limit of the \Sc functional. The
the tree-level contribution to $\log Z[g_{rs}]$
is thus the $\tau\rightarrow\infty$ limit of minus
the action evaluated on shell for a manifold with boundaries at 
$t=0$, where the induced metric is $g_{ij}$, and $t=\tau$ where it is
$g_{ij}+h_{ij}$. If we denote this as $W^{\rm tree}_{\tau,0}[g_{ij}+h_{ij},g_{ij}]$ then it satisfies the Hamilton-Jacobi equation, (which is the  tree-level \Sc equation).
This is simply the
statement that $W^{\rm tree}_{\tau+\delta\tau,0}$ can be obtained from
$W^{\rm tree}_{\tau,0}$ by allowing the fields to propagate according to the equations of motion from
$\tau$ to $\tau+\delta\tau$, i.e.

\be
W^{\rm tree}_{\tau+\delta\tau,0}[g_{ij}+h_{ij}+\delta\tau\dot h_{ij} ,g_{ij}]=
W^{\rm tree}_{\tau,0}[g_{ij}+h_{ij},g_{ij}]-L\,\delta\tau
\ee
where $L$ is the Lagrangian in (\ref{fiftwo}). Using
\be
-{\delta W^{\rm tree}_{\tau,0}\over\delta h_{ij}}={\sqrt g\over 2L\tau^{d-1}}\left(
\dot h^{ij}-g^{ij}\,\dot h^r_r\right)\equiv\pi_{ij}
\ee
this leads to

\bq
&&
 -{\partial W^{\rm tree}_{\tau,0}\over\partial\tau}
=L+\int d^dx\,\dot h_{ij}{\delta W^{\rm tree}_{\tau,0}\over\delta h_{ij}}
\nonumber\\
&&
\quad
=\int d^dx{\sqrt g L\over \tau^{d-1}}\left({d(d-1)\over L^2\tau^2}+R(g)\right)\nonumber\\
&&
+\int d^dx\left(-{\tau^{d-1}L\over \sqrt g} \pi^{ij}G_{ij rs} \pi^{rs}
+{\sqrt g L\over \tau^{d-1}}\left(h_{ij}\,\tilde G^{ij}(g)+
h_{ij}\,
\Box^{ijkl}h_{kl}\right)\right)+..\label{HJG}
\eq
with
\be 
G_{ij rs}=g_{ir}g_{js}-{1\over d-1}g_{ij}g_{rs}.
\ee
and the initial condition is
that $\exp W^{\rm tree}_{\tau,0}[\tilde g_{ij}, g_{rs}]\sim \delta[h_{ij}]$.
When the curvature tensors constructed from $g_{ij}$ are small,
and $h_{ij}$ is slowly varying we can expand $W^{\rm tree}$ in powers of $h$ 
and its derivatives as

\be  W^{\rm tree}_{\tau,0}[g_{ij}+h_{ij}, g_{rs}]
=\int {d^dx\,\sqrt g\over\tau^d}\left( -{d\over 4L}h_{ij}G^{-1\,ijrs}h_{rs}
+\Gamma_0+\Gamma_1^{ij}h_{ij}+h_{ij}\Gamma_2^{ijrs}h_{rs}+..\right)
\label{woo}
\ee
Apart from a constant term in $\Gamma_0$,
only the first term on the right has no derivatives of either $h_{ij}$ or
$g_{ij}$, and so
provides the dominant behaviour as $\tau\rightarrow 0$, satisfying the
initial condition (provided $h_i^i$ is suitably treated,
\cite{Hawking}). Substituting into (\ref{HJG}) and equating powers of $h_{ij}$
gives

\be
-{\partial \over\partial\tau}\left({\Gamma_0\over\tau^d}\right)
={1\over \tau^{d-1}}\left({d(d-1)\over L\tau^2}+LR(g)\right)
-{L\over\tau^{d+1}}\Gamma_1^{ij}G_{ij rs} \Gamma_1^{rs},
\label{freeg}
\ee
The free energy is the infinite $\tau$ limit of the $h_{ij}$
independent part of (\ref{woo}), i.e. $F[\tau, g_{ij}]=\int d^d x\,\sqrt g \,\Gamma_0/\tau^d$, and, as before, a Weyl scaling of the metric can be compensated by a scaling of $\tau$, so for $\delta g_{ij}=\delta\rho\,g_{ij}$ the change in $F$ is given by 

(\ref{freeg})

\be
\delta F=-{\delta\rho\over 2}\tau{\partial F\over\partial\tau}={\delta\rho\over 2}\int d^dx
\,\sqrt g L\left(
{1\over \tau^{d-2}}\left({d(d-1)\over L^2\tau^2}+R(g)\right)
-{1\over\tau^{d}}\Gamma_1^{ij}G_{ij rs} \Gamma_1^{rs}\right)
\label{freeg2}
\ee

Up to terms involving two derivatives, we have from (\ref{HJG})

\be
-{\partial \Gamma_1^{ij}\over\partial\tau}
=L\tau \tilde G^{ij}(g)\label{gam1}
\ee
Solving this for $\Gamma_1$ and substituting into (\ref{freeg2}) gives the variation as an expansion in powers of $\tau$ times the curvatures constructed from $g_{ij}$. We want to work
at finite $\tau$, so our expansion will be valid when we take the curvatures to
zero.

Now for ${\bf d=2}$ we have identically $\tilde G^{ij}=0$,
so $\Gamma_1=0$ and to the order we need (\ref{freeg2}) reduces to

\be
\delta F=\delta\rho\,\int d^2x
\,\sqrt g
{L}\left({1\over L^2\tau^2}+{1\over 2}R(g)\right)
\ee
from which we can identify the central charge as $c=24\pi L$, or,
since we have chosen units such that the three-dimensional 
gravitational constant satisfies $16\pi G_{\rm Newton}=1$, 
we have
$c=3L/(2G_{\rm Newton})$ as in \cite{Brown}. 

\medskip
\noindent
For ${\bf d=4}$ we obtain $\Gamma_1$ to the desired order from
(\ref{gam1}) as $\Gamma_1=-L\tau^2 \tilde G^{ij}/2$, so if we substitute into
(\ref{freeg}) we have

\be
\delta F={\delta\rho\over 2}\int d^4x
\,\sqrt g L\left(
\left({12\over L^2\tau^4}+{R(g)\over\tau^2}\right)
-L^2 \tilde G^{ij}G_{ij rs} \tilde G^{rs}/4\right)
\label{freego}
\ee
The first two terms represent divergences that should be cancelled by
counter-terms, the last piece is the finite Weyl anomaly. If we reinstate the 
five-dimensional Newton constant this becomes

\be
\delta\rho {L^3\over 128 \pi G_{\rm Newton}}\left( R_{ij}\,R^{ij}-R^2/3\right)
\ee
which agrees with \cite{Henningson}.

The results of this section can be generalized to higher-derivative gravity, as considered in \cite{odintsov}.

\vfill
\eject

\section{Conclusions}

We have interpreted the partition function for fields in Euclideanized anti de Sitter space-time as a limit of the \Sc functional. This allows canonical methods to be used, in contrast to the usual approach based on Green's functions. The former
have the advantage of separating out the time-dependence, and as a consequence we
have been able to compute one-loop effects in the AdS theory by solving the 
functional \Sc equation, (as well as reproducing known tree-level results
such as the scaling of two-point functions and the gravitational conformal anomaly). Although we have only studied the simplest one-loop
quantity, namely the anomalous scaling of the free energy for scalar and fermionic theories, these canonical methods can be expected to apply to more complicated
objects such as n-point functions, as they do in flat space \cite{PMJ}-\cite{Marcos}.

Our computation of the one-loop conformal anomalies for scalar and fermionic
theories shows that for generic values of the `mass' the anomalies
vanish, but for special values the anomalies have integer coefficients.
For example, when the boundary of the AdS spacetime is two-dimensional the 
Virasoro central charges are positive integers, $\tilde N$, when the mass
of the scalar field is $\sqrt{\tilde N^2-1}$  and when the mass of the
fermion is $\tilde N -1/2$. This implies that for generic values of the mass,
the boundary
CFTs for the scalar or fermi fields alone are non-unitary, since in a unitary theory a vanishing central charge implies an absence of quasi-primaries.
When the boundary is four dimensional the conditions
on the scalar and fermion masses for the conformal anomaly to be non-zero coincide with the mass spectra resulting from Kaluza-Klein compactification of Supergravity on $AdS_5\times S^5$, and when it is six dimensional they coincide with the mass spectra resulting from Kaluza-Klein compactification of Supergravity on $AdS_7\times S^4$. We expect that the same is true for the other fields
in the Supergravity theory, and this would make the one-loop corrections
to the tree-level calculation of \cite{Henningson}
that checks the Maldacena conjecture an intriguing prospect.

\bigskip
\bigskip
\noindent
{\bf Acknowledgements} PM would like to thank Roberto Emparan and Ivo Sachs
for useful conversations. DN is grateful to EPSRC for the award of a studentship.

\vfill\eject

\end{document}